\documentclass[manuscript]{aastex}
\def\tt{\mbox{\boldmath $\theta $}}
\def\pp{\mbox{\boldmath $\phi $}}
\shorttitle{COSMIC-RAY POSITRONS FROM MILLISECOND PULSARS}
\shortauthors{Venter et al.}

\begin{document}
\title{COSMIC-RAY POSITRONS FROM MILLISECOND PULSARS}
\author{C. VENTER\altaffilmark{1}, A. KOPP\altaffilmark{1,*}, A.~K. HARDING\altaffilmark{2}, P.~L. GONTHIER\altaffilmark{3}, AND  I.~B\"USCHING\altaffilmark{1}}
\altaffiltext{1}{Centre for Space Research, North-West University, Potchefstroom Campus, Private Bag X6001, Potchefstroom 2520, South Africa}
\altaffiltext{*}{On leave from Institut f\"ur Experimentelle und Angewandte Physik, Christian-Albrechts-Universit\"at zu Kiel, Leibnizstrasse~11, 24118 Kiel, Germany}
\altaffiltext{2}{Astrophysics Science Division, NASA Goddard Space Flight Center, Greenbelt, MD 20771, USA}
\altaffiltext{3}{Hope College, Department of Physics, Holland MI, USA}

\begin{abstract}
Observations by the \textit{Fermi} Large Area Telescope of $\gamma$-ray millisecond pulsar light curves imply copious pair production in their magnetospheres, and not exclusively in those of younger pulsars. Such pair cascades may be a primary source of Galactic electrons and positrons, contributing to the observed enhancement in positron flux above $\sim$10~GeV. \textit{Fermi} has also uncovered many new millisecond pulsars, impacting Galactic stellar population models. We investigate the contribution of Galactic millisecond pulsars to the flux of terrestrial cosmic-ray electrons and positrons. Our population synthesis code predicts the source properties of present-day millisecond pulsars. We simulate their pair spectra invoking an offset-dipole magnetic field. We also consider positrons and electrons that have been further accelerated to energies of several TeV by strong intrabinary shocks in black widow and redback systems. Since millisecond pulsars are not surrounded by pulsar wind nebulae or supernova shells, we assume that the pairs freely escape and undergo losses only in the intergalactic medium. We compute the transported pair spectra at Earth, following their diffusion and energy loss through the Galaxy. The predicted particle flux increases for non-zero offsets of the magnetic polar caps. Pair cascades from the magnetospheres of millisecond pulsars are only modest contributors around a few tens of GeV to the lepton fluxes measured by \textit{AMS$-$02}, \textit{PAMELA}, and \textit{Fermi}, after which this component cuts off. The contribution by black widows and redbacks may, however, reach levels of a few tens of percent at tens of TeV, depending on model parameters. 
\end{abstract}


\section{Introduction}
\label{sec:intro}
Recent measurements by \textit{PAMELA} \citep{Adriani09,Adriani13}, \textit{Fermi} Large Area Telescope \citep[LAT;][]{Ackermann12}, and the \textit{Alpha Magnetic Spectrometer} \citep[\textit{AMS$-$02};][]{Aguilar13,Aguilar14,Accardo14} have provided firm evidence that the positron fraction (PF) $\phi(e^+)/[\phi(e^+) + \phi(e^-)]$, with $\phi$ the flux, is increasing with energy above $\sim10$~GeV. Improved spectral measurements for 30 months of \textit{AMS$-$02} data extended the PF up to 500~GeV, and indicated a leveling off of this fraction with energy, as well as the PF being consistent with isotropy.

Secondary positrons are created during inelastic collisions between cosmic-ray nuclei and intergalactic hydrogen, which produce charged pions that in turn decay into positrons, electrons, and neutrinos. The fraction of this secondary component with respect to the total (electron + positron) cosmic-ray spectrum is expected to smoothly decrease with energy within the standard framework of cosmic-ray transport \citep[e.g.,][]{Moskalenko98}.\footnote{This, however, depends on model assumptions, i.e., a concave electron spectrum may lead to a rising PF with energy.} However, the \textit{AMS$-$02} electron spectrum is softer than the positron one in the range $20 - 200$~GeV \citep{Aguilar14}, and the measured PF rises with energy, pointing to nearby sources of primary positrons\footnote{Such an additional source of primary positrons may be either of dark matter annihilation origin \citep[e.g.,][]{Grasso09,Fan10,Porter11,Lin15}, or of astrophysical origin, including supernovae \citep[e.g.,][]{Blasi09,D10}, microquasar jets \citep{Gupta14}, molecular clouds \citep{Dogiel90}, pulsar wind nebulae \citep[e.g.,][]{Blasi11,Serpico12}, young or mature pulsars \citep[e.g.,][]{Arons81,HR87,Boulares89,Chi96,Zhang01,Grimani07,Profumo12,Kistler09,Hooper09,Yuksel09,Gendelev10,Yin13,Feng15}, `white dwarf pulsars' mainly formed by the merger of two white dwarfs \citep{Kashiyama11}, and millisecond pulsars \citep[MSPs;][]{Kisaka12}. In this paper, we investigate the latter source class.}. Moreover, the rising PF can be ascribed to a hardening of the positron spectrum (up to 200~GeV, after which it softens with energy), and not a softening in electron spectrum above 10~GeV. 

Alternatively, it has been argued that the observed rise in PF with energy may be explained purely by secondary positrons originating in the interstellar medium (ISM), without the need to invoke a primary positron source. \citet{Shaviv09} demonstrated that an inhomogeneous distribution of supernova remnants (SNRs), such as a strong concentration in the Galactic spiral arms, may explain the PF shape \citep[see also][who note that an unrealistically steep index for the primary electron spectrum needs to be invoked when assuming a homogeneous or smoothly varying source distribution; however, they do find evidence for an extra / secondary charge-symmetric electron-positron source to explain the data]{Gaggero14}. \citet{Moskalenko13} pointed out that the concave shape of the primary electron spectrum of \citet{Shaviv09} introduces an arguably artificial rise in the PF. \citet{Cowsik10} put forward a model assuming that a significant fraction of the boron below 10~GeV is generated through spallation of cosmic-ray nuclei in small regions around the sources. In this case, the contribution from spallation in the ISM would have a flat or weak energy dependence, and the GeV positrons would almost exclusively be generated through cosmic-ray interactions in the ISM. \citet{Moskalenko13} noted that such sources should be observable as very bright GeV $\gamma$-ray sources with soft spectra, while the diffuse emission would be significantly dimmer than observed. This scenario is also at odds with current estimates of the supernova birth rate. \citet{Blum13} found an upper bound to the positron flux by neglecting energy losses, arguing that the flattening of the PF seen by \textit{AMS-02} around several hundred GeV is consistent with a purely secondary origin for the positrons. \citet{Moskalenko13} noted that their arguments imply quite hard injection spectra for primary nuclei, in contradiction to $\gamma$-ray observations of SNRs that seem to require rather steep spectra. In addition, a very fast escape time for the positrons is implied, and if this is extrapolated to higher energies, it would lead to a large cosmic-ray anisotropy, which has not been observed. \citet{Dado15} furthermore conclude that if the energy losses of positrons in the ISM are included in the transport calculation, the upper limit to the positron flux is much lower than the limit derived by \citet{Blum13}, requiring a primary source of positrons in this case.

MSPs are the oldest population of rotation-powered pulsars, characterized by low surface magnetic fields, and are thought to have acquired their very short periods through spin-up by accretion from a binary companion \citep{Alpar82}. For the most part, they have not been considered as an important source of cosmic-ray positrons since the majority lie below the death lines for high-multiplicity pair cascades (assuming dipole magnetic fields; \citealt{HMZ02,ZC03}) and were thus considered to be pair-starved \citep{HUM05}. However, this picture changed with the detection of pulsed $\gamma$-ray emission from a large number of MSPs by {\it Fermi} \citep{Abdo13}. Most of the $\gamma$-ray light curves show narrow, double peaks trailing the radio peaks, very similar to those of younger pulsars. Such light curves can only be fit by outer magnetospheric gap models \citep{Venter09,Johnson14}. The existence of narrow accelerator gaps requires large numbers of electron-positron pairs (high multiplicity) to screen the electric field parallel to the magnetic field in the open magnetosphere interior to (at lower colatitudes than) the gaps. It has been suggested that distortions of the surface magnetic field may increase pair production for MSPs, either in the form of higher multipoles \citep[e.g.,][]{ZC03}, or offset polar caps \citep[PCs;][]{Arons96,HM11a,HM11b}. \citet{HM11a} found that even small offsets of the PC from the magnetic axis (a small fraction of the stellar radius) can greatly enhance the pair multiplicity. This is due to the increase in accelerating electric field on one side of the PC, which stems from the decrease in curvature radius of the distorted magnetic field. Furthermore, MSPs produce pairs with energies around 100 times higher that those of young pulsars, due to their relatively low magnetic fields which require a higher photon energy for magnetic photon pair production to take place. In this case, the pair spectra extend to several TeV \citep{HM11b}. There has also recently been a substantial increase in the population of known MSPs through discovery of new radio MSPs in {\it Fermi} unidentified sources \citep{Abdo13}. Many of these are nearby (within 1 kpc) and a number are relatively bright, indicating that the existing radio surveys were incomplete (or insensitive to the detection of many MSPs). All of the above factors \textit{(more sources characterized by higher pair multiplicities and larger maximal particle energies than previously thought)} make the study of MSPs as sources of cosmic-ray electrons and positions quite attractive.

We have previously studied the contribution to the terrestrial electron spectrum by the nearby MSP PSR~J0437$-$4715 assuming a pair-starved potential, but found the contribution of this nearby MSP to be negligible within this model. We also considered the contribution of the much younger Geminga \citep[see also][]{Aharonian95}, and found that it may contribute significantly, depending on model parameters \citep{Buesching08}. \citet{Buesching08b} furthermore noted that both Geminga and PSR B0656+14 may be dominant contributors to the terrestrial positron flux, and may be responsible for an anisotropy of up to a few percent in this flux component. 
We have recently made a first attempt to carefully assess the contribution of MSPs (excluding those found in globular clusters) to the cosmic-ray lepton spectrum at Earth \citep{Venter15}, where we have considered pairs originating in cascades within the magnetospheres of MSPs. However, since about 80\% of MSPs have binary companions, in some fraction of these systems shocks may form in the pulsar winds as they interact with the companion wind or atmosphere \citep{HG90,Arons93}, which could accelerate the pairs to higher energies. It is possible that such shock acceleration occurs in some black widow (BW) systems, such as PSR B1959+20 \citep{Arons93}. Due to \textit{Fermi} observations, the population of BWs and redbacks (RBs) has increased significantly. We therefore now also study the effect of pairs that have been reaccelerated in intrabinary shocks of BW and RB systems. We furthermore include Klein-Nisihna (KN) effects \citep{Ruppel10,Blies12} when assessing the inverse Compton (IC) loss rate the particles suffer as they traverse the interstellar radiation field (ISRF) of the Galaxy. We describe the  assumed source properties of MSPs by first discussing the central expectation of roughly equal numbers of electrons and positrons coming from pulsar magnetospheres (Section~\ref{sec:pair_prod}), after which we describe our population synthesis code used to predict the present-day number of MSPs as well as their location and power (Section~\ref{sec:synth}). We describe an additional BW / RB source population in Section~\ref{sec:BW}. Moving to source spectra, we describe our PC pair cascade code that yields realistic pair spectra (Section~\ref{sec:pairs}). We also describe the spectra injected by BW and RB systems (Section~\ref{sec:BW2}), and motivate why we neglect the small contribution due to primaries (Section~\ref{sec:prim}). We next discuss our assumptions regarding the ISRF (Section~\ref{sec:ISRF}) and Galactic magnetic field strength (Section~\ref{sec:Bfield}), which are necessary inputs to the calculation of energy losses suffered by the leptons (Section~\ref{sec:losses}). We use this together with a prescription for particle diffusion when solving a transport equation (Section~\ref{sec:transport}) to calculate the spectra at Earth (Section~\ref{sec:results}). We discuss our results in Section~\ref{sec:discuss}, while our conclusions follow in Section~\ref{sec:concl}. 


\section{Millisecond pulsars as sources of cosmic-ray electrons and positrons}
We first address the question of pair production in pulsar magnetospheres (Section~\ref{sec:pair_prod}), specifically as this pertains to MSPs, before describing two pulsar populations we consider in the rest of the paper: Galactic MSPs resulting from population synthesis modeling (Section~\ref{sec:synth}), and BWs and RBs which may further accelerate particles flowing out of the MSP magnetospheres in their intrabinary shocks. (Section~\ref{sec:BW}).

\subsection{Pair production in pulsar magnetospheres}
\label{sec:pair_prod}

Production of electron-positron pairs in pulsar magnetospheres, first proposed by \citet{Sturrock71}, is widely considered to be critical for supplying charges to the magnetosphere as well as plasma for the observed coherent radio emission. The pairs can be efficiently produced in electromagnetic cascades above the PCs \citep{DH82} by $\gamma$ rays that undergo conversion to electron-positron pairs by the strong magnetic field \citep{Erber66}. These cascades are initiated by the acceleration of primary electrons in strong electric fields above the neutron star surface. Curvature and IC radiation from these particles reaches tens of GeV, creating pairs in excited Landau states. The pairs lose their perpendicular momentum by emitting synchrotron radiation (SR) photons that create more pairs. In young pulsars with magnetic fields above $10^{12}$ G, the cascades can produce multiplicities of $10^3 - 10^4$ pairs per primary electron \citep{DH82,HM11a}. The dense pair plasma will screen the accelerating electric field above the gap, except in a narrow gap along the last open field lines \citep{MH04}. Screening by pairs may provide nearly force-free conditions \citep[e.g.,][]{Spitkovsky06} throughout the magnetosphere, maintaining the narrow accelerator and emission gaps necessary to produce the sharp caustic $\gamma$-ray peaks observed by \textit{Fermi}. The pair plasma created by pulsars flows out of the magnetosphere along open magnetic field lines close to the pole and provides the radiating particles for the surrounding PWNe.  Models of PWNe require high pair multiplicity to produce the observed SR and IC emission \citep{DeJager96,Bucc11}.  

Most MSPs, because of their very low magnetic fields, have difficulty producing high-multiplicity pair cascades initiated by curvature radiation if the surface fields are dipolar. They are able to produce cascades from IC radiation, but these cascades do not have high enough multiplicity to screen the electric fields \citep{HMZ02}.They were thus assumed to have pair-starved magnetospheres \citep{HUM05} that have particle acceleration on all open field lines up to high altitudes. Such magnetospheres would produce broad $\gamma$-ray peaks \citep{Venter09} at earlier phase than the radio peak. However, \textit{Fermi} detected MSPs with narrow peaks in their $\gamma$-ray light curves arriving at later phase than the radio peak, very similar to those of young pulsars, implying that MSPs are somehow able to produce the high multiplicity pair cascades required to screen most of the open field region. \citet{HM11a,HM11b} suggested that MSPs have non-dipolar fields near their surface that enhance the accelerating electric fields and enable creation of more pairs. Introducing a generic toroidal component to the dipole field that effectively caused an offset of the PC relative to the magnetic pole, \citet{HM11a} were able to specify the field distortion with two offset parameters, $\varepsilon$ and $\phi_0$, describing the magnitude and azimuthal direction of the shift. Physically, $\varepsilon \sim 0.1$ for MSPs corresponds to the PC offset caused by the sweepback near the light cylinder of a vacuum retarded dipole field \citep{Deutsch55,Dyks04}, $\varepsilon \sim 0.2$ to the PC offset from sweepback of a force-free field \citep{Spitkovsky06}, and $\varepsilon > 0.2$ to the PC offset by multiple fields near the surface. \citet{HM11b} found that for magnetic fields with $\varepsilon > 0.4$, requiring moderate surface multipole components, all known MSPs were able the produce pair cascades by curvature radiation.  

Aside from the requirement of field distortions to produce higher pair multiplicity for the $\gamma$-ray profiles, there is evidence of a non-dipolar surface field structure in MSPs from the study of their X-ray emission. The thermal X-ray pulse profiles of some MSPs show asymmetries that require offsets from the magnetic axis of the emitting hot spot on the neutron star surface in order to successfully fit the light curves.  Since the emission likely originates from PC heating, it is argued that MSPs such as PSR~J0437$-$4715 \citep{Bogdanov07,Bogdanov13} and PSR~J0030+0451 \citep{BogdanovGrindlay09} have either offset dipoles or offset PCs.  The shift of the heated PC needed for modeling the light curve of PSR~J0437$-$4715, $\sim 2$ km, corresponds to an offset parameter $\varepsilon \sim 0.6$. (In what follows, we will adopt values of $\varepsilon=0.0, 0.2,$ and 0.6 in our modeling.)

Below, we discuss two classes of MSPs which we consider to be sources of cosmic-ray electrons and positrons.

\subsection{Galactic synthesis model for the present-day MSP population}
\label{sec:synth}
We implement the results of a new study by \citet{Gonthier15} of the population synthesis of radio and $\gamma$-ray MSPs that lead to the present-day distribution of MSPs. This is assumed to be an equilibrated distribution within the Galaxy whose evolution has been described in Section~3 of the work of \citet[][hereafter SGH]{Story07} where the radial ($\rho$ in cylindrical coordinates) distribution was assumed to be that of \citet{Pacz90}, with a radial scaling of 4.5~kpc and a scale height of 200~pc, instead of 75~pc used in that work. In addition, the supernova kick velocity model that was implemented was that of \citet{Hobbs05} using a Maxwellian distribution with a width of 70~km\,s$^{-1}$ (resulting in an average speed of 110~km\,s$^{-1}$). The Galaxy is seeded with MSPs treated as point particles with ages going back to the past 12~Gyr assuming a constant birth rate of $4.5\times10^{-4}$ MSPs per century as obtained in SGH. The MSPs are evolved in the Galactic potential from their birth location to the present time when an equilibrium distribution has been established. 

We assume that MSPs are ``born'' on the spin-up line with initial period $P_0$ dependent on the surface magnetic field $B_{\rm s}$, which we assume does not decay with time. We assume a power-law distribution for the magnetic fields. As in the case of the study of SGH, the simulation prefers a power-law distribution of periods $P_0(B_8)$, with an index of $\alpha_{\rm B}$, with a normalized distribution given by the expression
\begin{equation}
P_0\left(B_8\right) = \frac{(\alpha+1)\,B_8^{\alpha_{\rm B}}}{ B_{\rm max}^{\alpha_{\rm B}+1}-B_{\rm min}^{\alpha_{\rm B}+1}},\label{eq:Bf}
\end{equation}
where $B_8 = B_{\rm s}/(10^8$~G) and $B_{\rm max} = 10^3$. We consider $\alpha_{\rm B}$ and $B_{\rm min}$ to be free parameters, which are then fixed at optimum values. In the study of SGH a preferred index of $-1$ was used. However, improved agreement with the new simulation is achieved with an index of $\alpha_{\rm B}=-1.3$.

We assume a distribution of mass accretion birth lines, from the Eddington critical mass accretion rate to about $10^{-3}$ of the critical value, following the study by \citet{Lamb2005}. We parameterize the mass accretion rates with a line in the $\dot P-P$ diagram as was done in Equation~(5) of SGH. The intercept of this birth line was dithered using a dithering parameter $\delta$. The study of SGH used a ramp distribution of $\delta$ characterized by a linear function increasing with $\delta$. We found improved agreement by uniformly dithering $\delta$ between 0 and 2, with the restriction that the birth period $P_0\,>\, 1.3\ {\rm ms}$.

Recently, significant progress has been made in obtaining more realistic pulsar magnetosphere solutions than the retarded, vacuum dipole \citep{Deutsch55}. Force-free electrodynamic solutions were obtained by \citet{Spitkovsky06} leading to the following prescription for the pulsar spin-down power
\begin{equation}
L_{\rm sd} \sim \frac{2\,\mu^2\, \Omega^4}{3\,c^3}\left(1+\sin^2\alpha\right),\label{eq:Spit}
\end{equation}
where $\mu$ is the magnetic dipole moment, $\Omega$ is the rotational angular velocity, $c$ is the speed of light, and $\alpha$ is the magnetic inclination angle relative to the pulsar's rotational axis.  Considering accelerating fields and force-free solutions, \citet{Li12} constructed solutions of magnetospheres filled with resistive plasma, arriving at a very similar spin-down formula. \citet{Contopoulos14} considered the ideal force-free magnetosphere everywhere except within an equatorial current layer, and also arrived at a similar prescription for $L_{\rm sd}$. These results encourage us to implement such a spin-down model into our population synthesis code. Using a dipole moment of $\mu=B_{\rm s}\,R^3\,/\, 2$, where $R$ is the stellar radius and $B_{\rm s}$ the surface field at the pole, and equating $L_{\rm sd}$ to the rate of rotational energy loss yields the expression
\begin{equation}
B^2_{\rm s} = \frac{6\,c^3 I P \dot P}{4 \pi^2\,R^6\left(1+\sin^2\alpha\right)}.\label{eq:Spin} 
\end{equation}
Integrating this equation over the age $t$ of the pulsar provides the expression for obtaining the present-day period~$P(t)$ 
\begin{equation}
P^2 =P_0^2 +  \frac{ 4 \pi^2\,R^6}{3\,c^3 I } \left(1+\sin^2\alpha\right)\,B_{\rm s}^2\, t.\label{eq:PresP}
\end{equation}
We assume $R\,=\,12\ {\rm km}$ and MSP mass $M_{\rm MSP}=\,1.6\ M_\odot$, where $M_\odot$ is the mass of the Sun. We use the prescription outlined in Section~2 of \citet{Pierbattista12} to obtain the moment of inertia, which with these values of $R$ and $M_{\rm MSP}$ yields a value of $I = 1.7 \times 10^{45}\  {\rm g\,cm^2}$. While there is growing evidence that the inclination angle becomes aligned with the neutron star's rotational axis with time in the case of normal pulsars \citep{Johnston07,Young10}, we do not consider such an alignment model in the case of MSPs.

\begin{figure}
\epsscale{1.0}
\plotone{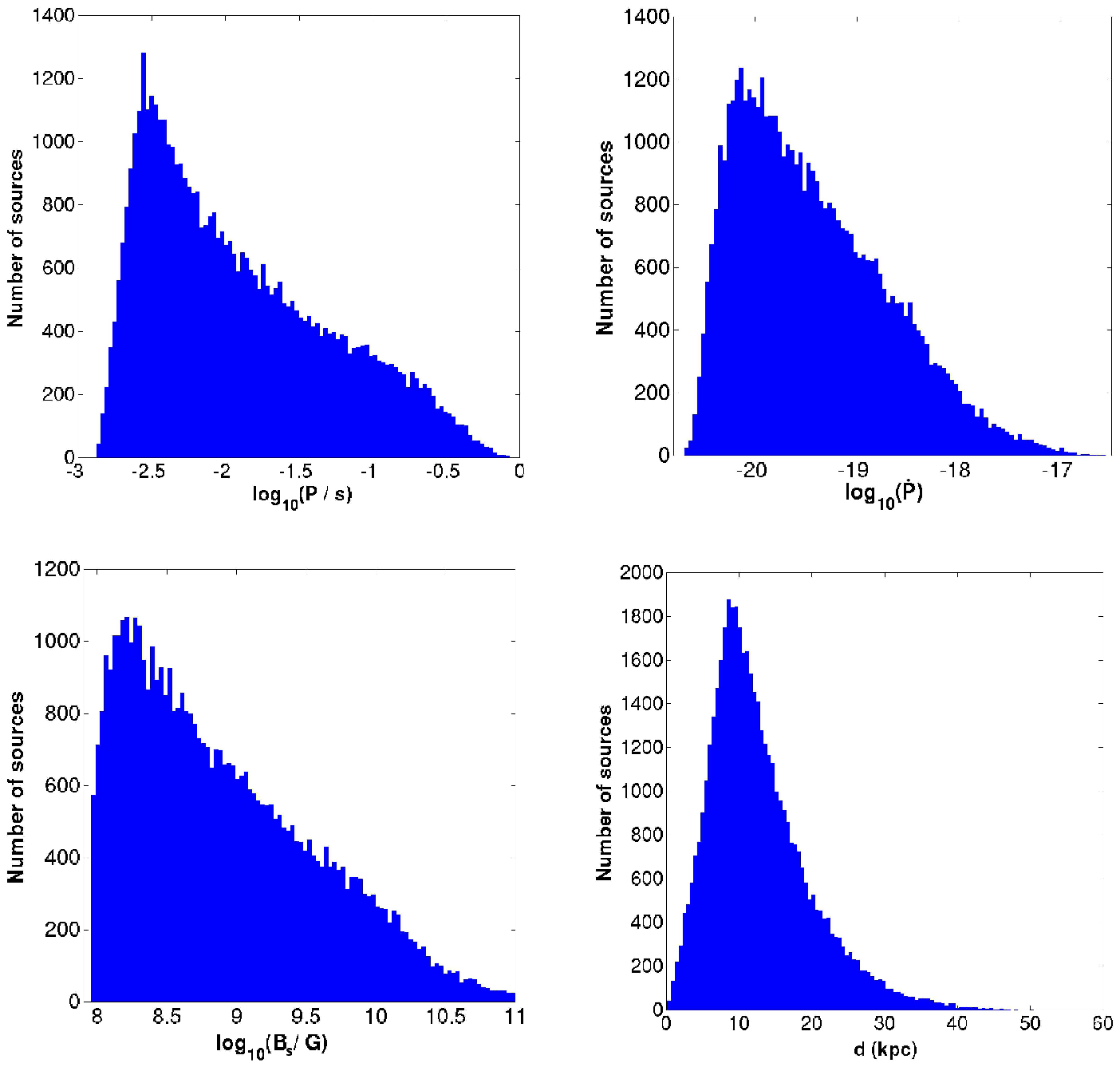}
\caption{Histograms of period $\log_{10}(P)$, period derivative $\log_{10}(\dot{P})$, surface magnetic field $\log_{10}(B_{\rm s})$, and distance $d$ characterizing the simulated present-day Galactic MSP population (Section~\ref{sec:synth}).}\label{fig:synth1}
\end{figure}

\begin{figure}
\epsscale{1.0}
\plotone{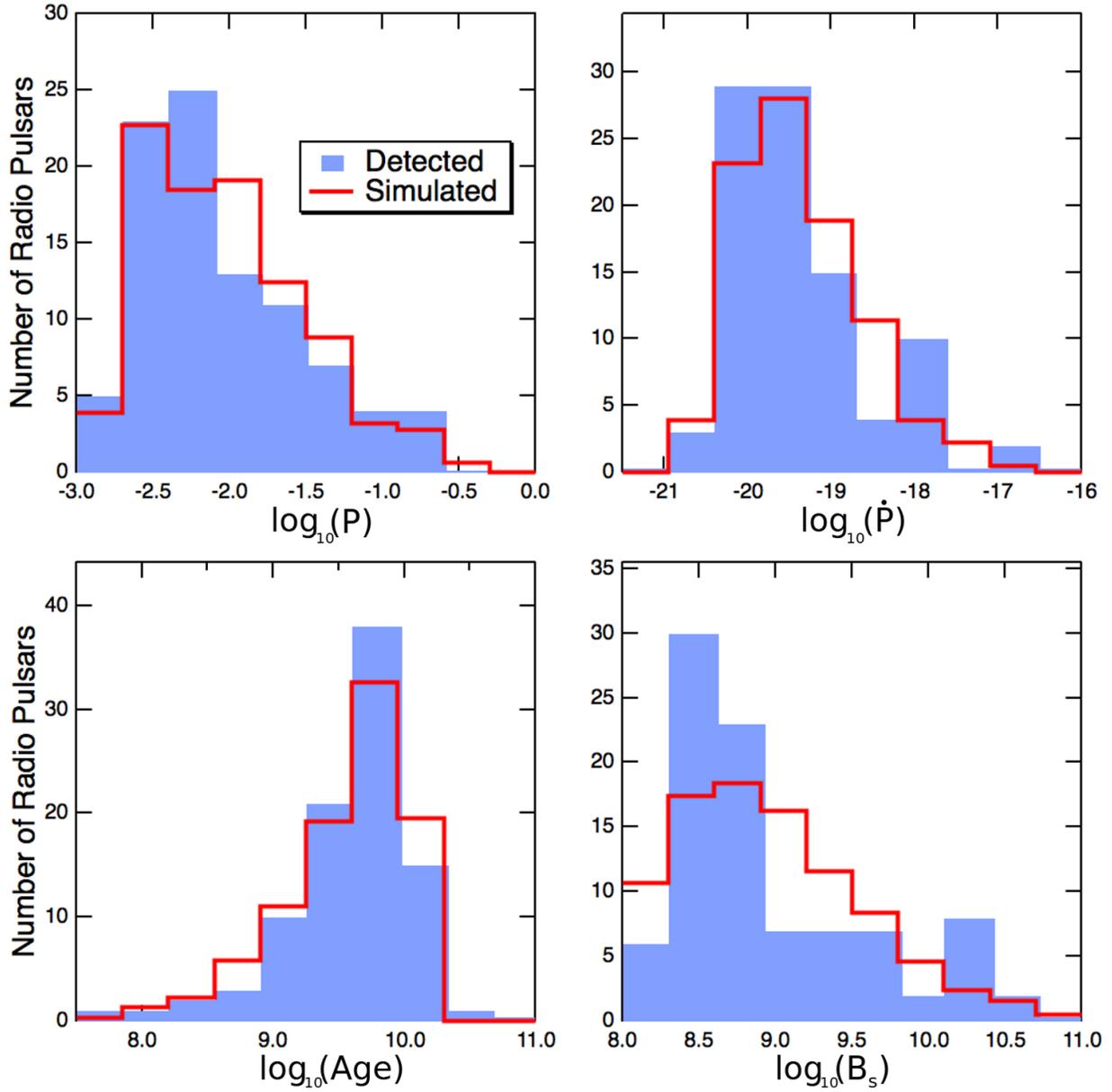}
\caption{Comparison of several simulated and measured properties of a population of detected radio-loud MSPs. Adapted from \citet{Gonthier15}.}\label{fig:synth2}
\end{figure}

Figure~\ref{fig:synth1} indicates histograms of period $\log_{10}(P)$, period derivative $\log_{10}(\dot{P})$, surface magnetic field $\log_{10}(B_{\rm s})$, and distance $d$ characterizing the simulated present-day Galactic MSP population. Figure~\ref{fig:synth2} shows several best-fit simulated and observed radio properties ($\log_{10}(P)$, $\log_{10}(\dot{P})$, characteristic age $\log_{10}(\tau_{\rm c})$, and $\log_{10}(B_{\rm s})$) of radio-loud MSPs detected in 12 radio surveys. The output from this simulation predicts the location as well as $P$ and $\dot{P}$ of roughly 50,000 Galactic MSPs, which we use as discrete sources of relativistic electrons and positrons in the calculations that follow.

\subsection{MSPs in binary systems -- BWs and RBs}
\label{sec:BW}
\begin{figure}
\epsscale{1.0}
\plotone{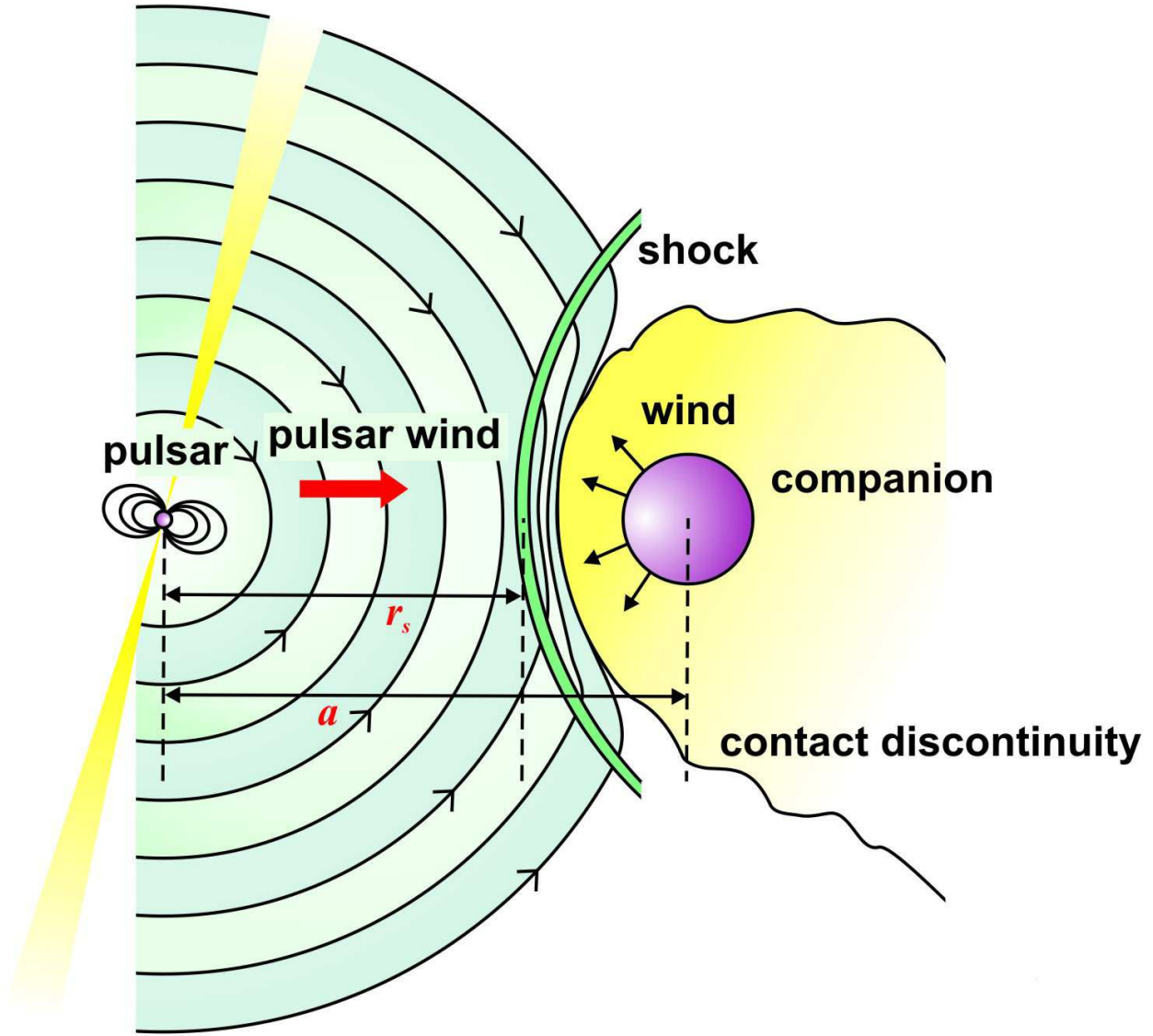}
\caption{Schematic of the formation of a shock upon collision of pulsar and companion winds. Adapted from \citet{Harding90}.}\label{fig:shock_cartoon}
\end{figure}

The majority of MSPs (about 80\%) are in binary systems, and a subset of these, the BWs and RBs, may contain strong intrabinary shocks that can further accelerate the pairs. BWs are close binary systems, with orbital periods of hours, containing a rotation-powered MSP and a compact companion having very low mass, $\sim 0.01 - 0.05\, M_{\sun}$. The companion stars in BWs undergo intense heating of their atmospheres by the MSP wind, which drives a stellar wind and rapid mass loss from  the star. A shock will form in the pulsar wind at the pressure balance point of the two winds and particle acceleration may occur in these shocks \citep{HG90,Arons93}. RBs are similar systems, except that the companions have somewhat higher masses, $\sim 0.1 - 0.4\, M_{\sun}$ \citep{Roberts11}.
The MSPs in both types of system are typically energetic, with $L_{\rm sd} \sim 10^{34} - 10^{35}\,\rm erg\,s^{-1}$. Figure~\ref{fig:shock_cartoon} is a schematic view of a shock formed between the colliding pulsar and companion star winds.

Before the launch of \textit{Fermi} these systems were rare, with only three BWs and one RB known.  The large amount of material blown off from the companion stars absorbs and scatters the radio pulsations from the MSPs, making them difficult to detect at radio wavelengths.  In the last few years, radio searches of \textit{Fermi} unidentified $\gamma$-ray point sources \citep{Ray12} have discovered 14 new BWs and 6 new RBs to date, making a present total of 24 of these systems.
In order to assess the contribution of these systems to the Galactic cosmic-ray positrons, we compiled a list of public detections, plus some measured and derived quantities (see Tables~\ref{tab1} and~\ref{tab2}). In deriving the spin-down luminosity and surface magnetic fields for the pulsars in these systems, we used an MSP radius of $R=9.9\times10^5$~cm and moment of inertia of $1.56\times10^{45}$~g\,cm$^2$, in order to be consistent with our pair cascade model assumptions (Section~\ref{sec:pairs}).

Evolution models and population synthesis of MSP binary systems yield a birthrate for BW systems $\sim 1.3 \times 10^{-7}\,\rm yr^{-1}$ \citep{Kiel13}. Taking an age of the Galaxy around 12 billion years, there may be a total population of several thousand BW systems. Since only a small fraction of these have been discovered, it is harder to estimate how many undiscovered BW and RB systems are within several kpc of Earth. Conservatively, the known nearby population may be $\sim$ 10\% of the total, or around several hundred. By considering only the 24 known BWs and RBs, we are obtaining a lower limit to the cosmic-ray flux contribution by binary MSPs.


\section{Models for pair injection spectra}

\subsection{Computation of pair spectra from pulsar polar caps}
\label{sec:pairs}
We calculate the spectra of pairs leaving the MSP magnetosphere using a code that follows the development of a PC electron-positron pair cascade in the pulsar magnetosphere \citep[details of the calculation can be found in][]{HM11b}. The pair cascade is initiated by curvature radiation of electrons accelerated above the PCs by a parallel electric field, derived assuming space-charge-limited flow (i.e., free emission of particles from the neutron star surface; \citealt{AS79}). A fraction of the curvature photons undergo magnetic pair attenuation \citep{Erber66,DH83}, producing a first-generation pair spectrum which then radiates SR photons that produce further generations of pairs. The total cascade multiplicity $M_+$ (average number of pairs spawned by each primary lepton) is a strong function of pulsar period $P$ and surface magnetic field strength $B_{\rm s}$, so that many pulsars with low magnetic fields and long periods produce either few or no pairs for dipole field structure ($\varepsilon = 0$), leading to a pair death line in the $P\dot{P}$ diagram.

\begin{figure}
\epsscale{1.0}
\plotone{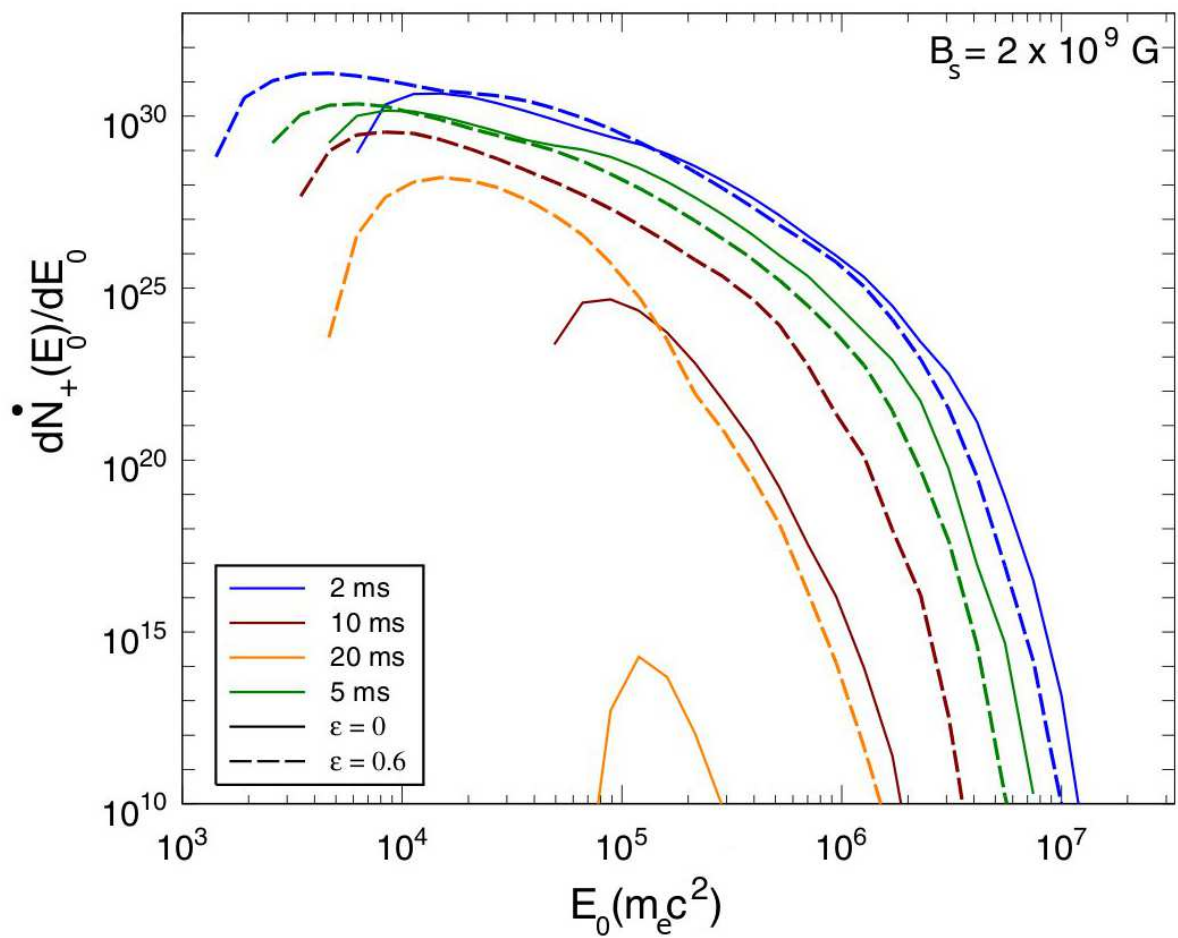}
\caption{Sample electron-positron pair spectra (number of pairs per second and energy) calculated for different periods $P$ and offset parameters $\varepsilon$, as indicated in the legend, and for a fixed $B_8 = 20$ (i.e., $B_{\rm s}=2\times10^9$~G). The $x$-axis indicates source energy in units of $m_ec^2$. From \citet{HM11b}.}\label{fig:pairspectra}
\end{figure}

However, as discussed in Section \ref{sec:pair_prod}, the sweepback of magnetic field lines near the light cylinder (where the corotation speed equals the speed of light) as well as asymmetric currents within the neutron star may cause the magnetic PCs to be offset from the dipole axis. We adopt the distorted magnetic field structure introduced by \citet{HM11b} that leads to enhanced local electric fields, boosting pair formation, even for pulsars below the pair death line. \citet{HM11b} considered two configurations for the dipole offset in which the magnetic field is either symmetric or asymmetric with respect to the dipole axis. Sweepback of the global field would produce asymmetric offsets, while the observed offset in the MSP J0437$-$4715 is symmetric \citep{Bogdanov13}. We adopt a symmetric field structure for calculating the pair spectra of MSPs in this paper. In the symmetric case, the magnetic field in spherical polar coordinates ($\eta $, $\theta $, $\phi $) is
\noindent 
\begin{equation}
{\bf B} \approx {B_{\rm s}\over {\eta ^3}}~\left[ \hat{\bf r}~\cos \theta +{1\over 2} ~\hat{\tt}~ (1+a)~\sin \theta - \hat{\pp }~\varepsilon ~\sin \theta~\cos \theta~\sin (\phi - \phi_0) \right],
\label{B1}
\end{equation}
where $B_{\rm s}$ is the surface magnetic field strength at the magnetic pole, $\eta = r/R$ is the dimensionless radial coordinate in units of neutron star radius $R$, $a=\varepsilon ~\cos (\phi - \phi _0)$ is the parameter characterizing the distortion of polar field lines, and $\phi_0$ is the magnetic azimuthal angle defining the meridional plane of the offset PC.  Using this field structure, \citet{HM11b} derive the component of the electric field parallel to the local magnetic field, $E_\parallel$, that accelerates electrons. We have used the $E_\parallel$ of Equation~(11) of \citet{HM11b} that corresponds to a symmetric offset and use these field structures to accelerate the electrons above the PC to simulate the pair cascades. The pair spectra (Figure~\ref{fig:pairspectra}) are characterized by $P$, $\dot{P}$ (or equivalently, $B_{\rm s}$ via Equation~[\ref{eq:Spin}]), and offset parameter $\varepsilon$. From our simulations, we find that about $\sim1$\% of $L_{\rm sd}$ is tapped to generate the pairs.

We used a grid in $P$ and $B_{\rm s}$ encompassing $P = \left(1, 1.8, 2, 2.5, 3, 4, 5, 7, 10, 20, 50, 100\right)$~ms, and $B_8 = \left(1, 1.5, 2, 3, 5, 8, 10, 15, 20, 50\right)$. For each source in the present-day MSP population with predicted values of $P$ and $\dot{P}$ (Section~\ref{sec:synth}), we found its associated pair spectrum by interpolating spectra on this grid. We used an inclination angle of $\alpha=45^\circ$, mass $M_{\rm MSP} = 2.15M_\odot$, radius $R = 9.9$~km, and moment of inertia $I = 1.56\times10^{45}$~g\,cm$^2$ for all MSPs. We adopted an equation of state with larger $M_{\rm MSP}$ here (and associated smaller $I$) compared to that used in the population code (Section~\ref{sec:synth}), since some MSPs have measured masses $M_{\rm MSP}\sim 2 M_\odot$ \citep{Demorest10}, and this enhances pair multiplicity. However, this discrepancy is removed by considering a large range of $\varepsilon$, since the latter simulates a large range of pair multiplicities that would correspond to different equations of state, and thus different values of $M_{\rm MSP}$. We used dipole offsets of $\varepsilon = (0.0,0.2,0.6)$ and set $\phi_0=\pi/2$ (this parameter controls the direction of offset of the PC).  

We use the above spectra as input for the calculation of the positron component from the population-synthesis sources (Sections~\ref{sec:synth} and~\ref{sec:results}). Since MSPs are not surrounded by nebulae that can trap the pairs and degrade their energy before escape, we can assume that the pair spectra emerging from the MSPs are good representations of the actual source spectra. 

\subsection{Spectra from particles accelerated in the intrabinary shocks of BWs and RBs}
\label{sec:BW2}

We assume that the pairs escaping from the pulsar magnetosphere may be further accelerated in the intrabinary shock that originates between the pulsar and companion winds in BW and RB systems. Acceleration of leptons at a large distance outside the pulsar light cylinder is necessary to account for the extended SR emission observed from PWNe. Such acceleration is thought to occur at or near the termination shock in the pulsar wind \citep{KC84} that is confined by the sub-relativistic expansion of the surrounding supernova shell. The acceleration mechanism near the pulsar wind termination shocks is not understood, but is known to be highly efficient, since the bolometric luminosity of the Crab nebula is about 20\% of the pulsar spin-down luminosity and the inferred maximum particle energy, $\sim 10^{16}$ eV, is at least 10\% of the available voltage across open field lines \citep{DeJager96}. The pulsar wind termination shock is relativistic and perpendicular, so that the diffusive first-order Fermi mechanism becomes problematic unless most of the magnetic energy is converted into particle energy upstream of the shock \citep{Sironi11a}. However, either shock-driven reconnection \citep{Sironi11b} or strong electromagnetic waves \citep{AK13} could cause demagnetization, enabling diffusive acceleration to proceed.  

Regardless of the acceleration mechanism, the maximum particle energy will be limited by the universal scaling, $E_{\rm max} \sim vBR_s/c$ \citep{Harding90}, where $v$ is a bulk flow velocity, $B$ is the magnetic field strength, and $R_s$ is a scale size of the system.  In the case of shock acceleration, the maximum energy comes from a balance between the minimum acceleration timescale, set by the particle diffusion, and the timescale for escape from the shock of radius $R_s$. However, for leptons, the timescale for SR losses is shorter than the escape time and the maximum energy will be set by balancing the acceleration timescale with the SR loss timescale.

We assume that the reaccelerated shock-accelerated spectrum will be an exponentially cut off power law with spectral index of $-2$
\begin{equation}
  Q_i(E) = Q_{0,i}E_0^{-2}\exp\left(-\frac{E_0}{E_{\rm cut}}\right),\label{eq:bw_spec}
\end{equation}
with the index $i$ indicating the $i^{\rm th}$ source, $Q_{0,i}$ the normalization factor, and $E_0$ the particle energy at the source position. In order to estimate the maximum (cutoff) energy, we balance the energy gain rate from shock acceleration, assuming Bohm diffusion, and the SR loss rate that particles experience in the strong magnetic field at the shock radius. This leads to the following expression \citep{HG90}
\begin{equation}
  E_{\rm cut} \approx 2.6 B_8^{-1/2}P_{\rm ms} a_{11}^{-1/2}\left[\frac{3\left(\xi-1\right)}{\xi\left(\xi+1\right)}\right]^{1/2}~{\rm TeV},
\end{equation}
with $P_{\rm ms}$ the pulsar period in milliseconds, $a_{11} = a/(10^{11}$~cm) the binary separation, and $\xi$ the shock compression ratio. This is slightly different from Equation~(34) in \citet{HG90}, since they assumed that the shock distance from the pulsar is $r_{\rm s} \approx a - R_*$, with $R_*$ the companion radius. For BWs and RBs, the shock is close to the companion star, and we assume $r_{\rm s} \approx a$, leading to the modified expression given above. The binary separation may be found as follows (given the extremely small eccentricities of these systems)
\begin{equation}
  a = \left[\frac{G\left(M_{\rm MSP} + M_{\rm comp}\right)}{4\pi^2}\right]^{1/3}P_{\rm b}^{2/3},
\end{equation}
with $M_{\rm MSP}$ the MSP mass, $M_{\rm comp}$ the companion mass, $P_{\rm b}$ the binary period, and $G$ the gravitational constant. We have listed the inferred values of $a_{11}$ and $E_{\rm cut}$ for each of the detected BWs and RBs in Tables~\ref{tab1} and~\ref{tab2}. 

Now, we can normalize the spectrum (e.g., \citealt{Buesching08b}) using
\begin{eqnarray}
  \int_{E_{\rm min}}^\infty Q_i\,dE_0 & = & \left[M_+(P,B_{\rm s},\varepsilon) + 1\right]\dot{n}_{\rm GJ}(P,B_{\rm s},\varepsilon)\label{eq:current}\\
  \int_{E_{\rm min}}^\infty Q_iE_0\,dE_0 & = & \eta_{\rm p,max}L_{\rm sd},\label{eq:power}
\end{eqnarray}
with $M_+$ the pair multiplicity, $\eta_{\rm p,max}$ the efficiency of conversion of spin-down power $L_{\rm sd}=4\pi^2I\dot{P}P^{-3}$ to particle power (or shock efficiency), and $\dot{n}_{\rm GJ}(P,B_{\rm s},\varepsilon)$ the Goldreich-Julian particle outflow rate, appropriate for offset-dipole fields (see Equation~[3] of \citealt{HM11b}) characterized by an offset parameter $\varepsilon$ (Section~\ref{sec:pairs}). The latter is similar to the classical expression \citep{Goldreich69}
\begin{equation}
  \dot{n}_{\rm GJ} = \frac{2cA_{\rm PC}\rho_{\rm GJ}}{e} = \frac{4\pi^2B_{\rm s}R^3}{2ceP^2},
\end{equation}
with $A_{\rm PC}$ the area of one PC, and $\rho_{\rm GJ}$ the Goldreich-Julian charge density. In Equation~(\ref{eq:current}), we therefore normalize the spectrum to the total (primary plus secondary) current. We found $M_+$ by interpolating values on a grid of $P$ and $B_{\rm s}$, while we calculated $\dot{n}_{\rm GJ}(\varepsilon)$ directly from the cascade code \citep{HM11b}.

\begin{figure}
\epsscale{1.0}
\plotone{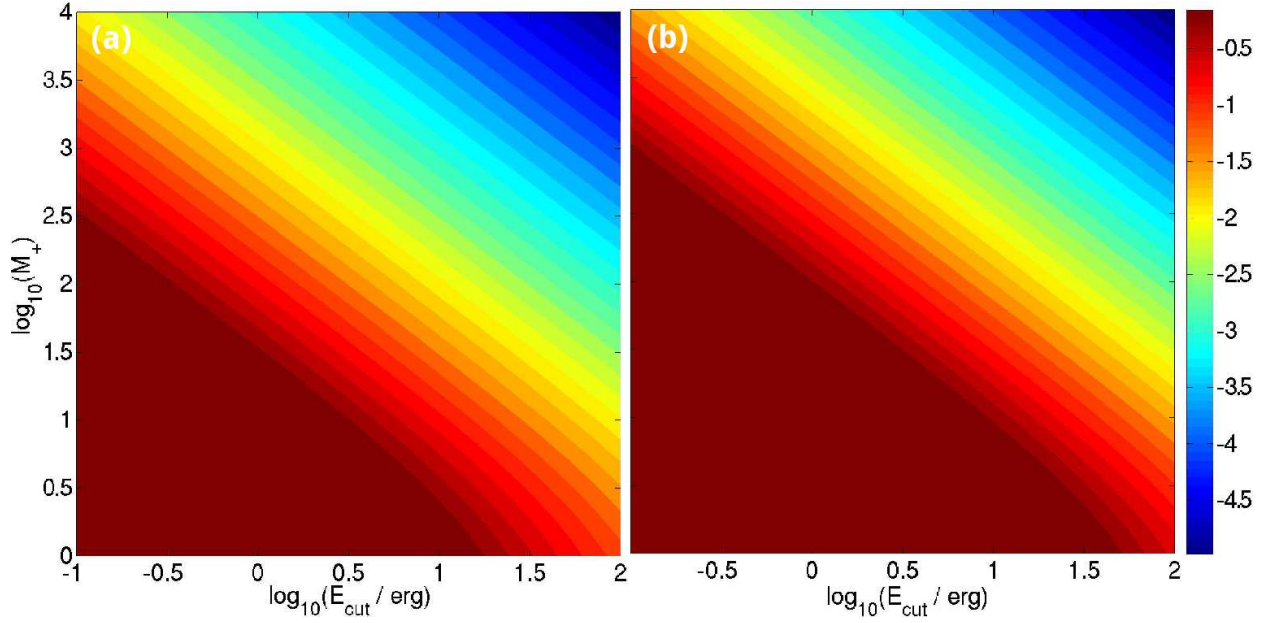}
\caption{Contour plot of $\log_{10}\left(E_{\rm min}/E_{\rm cut}\right)$ vs.\ $\log_{10}(M_+)$ and $\log_{10}(E_{\rm cut})$ assuming $P_{\rm ms}=3$ and $B_8 = 5$. Panel~(a) is for $\eta_{\rm p,max}=0.1$, and panel~(b) is for $\eta_{\rm p,max}=0.3$. Values near unity (dark red, i.e., the lower left corners) indicate that no solution could be found.}\label{fig:norm}
\end{figure}

The above is a system of two equations and two unknowns, $Q_{0,i}$ and $E_{\rm min}$ (when fixing $\eta_{\rm p,max}$). We find that the spectrum of Equation~(\ref{eq:bw_spec}) can only be normalized for some choices of $M_+$, $E_{\rm cut}$, and $\eta_{\rm p, max}$. Figure~\ref{fig:norm} shows contour plots of $\log_{10}\left(E_{\rm min}/E_{\rm cut}\right)$ vs.\ $\log_{10}(M_+)$ and $\log_{10}(E_{\rm cut})$ assuming $P_{\rm ms}=3$, $B_8 = 5$, $R=9.9\times10^5$~cm, and $I = 1.56\times10^{45}$~g\,cm$^2$. Panel~(a) is for $\eta_{\rm p,max} = 0.1$, while panel~(b) is for $\eta_{\rm p,max} = 0.3$. Values near unity (dark red regions, i.e., the lower left corners) indicate that no solution could be found for the given parameters. Fixing $\eta_{\rm p,max}$, one can see that for a fixed value of $E_{\rm cut}$, some minimum value of $M_+$ is required in order to find a physical solution $E_{\rm min} < E_{\rm cut}$. This is because a higher $M_+$ will raise $Q_{0,i}$, allowing Equation~(\ref{eq:current}) to be satisfied. A higher value of $E_{\rm min}$ has the same effect. For an even higher value of $M_+$ (typically associated with a higher value for $\varepsilon$) than the critical one needed to find a physical solution, the constraint on $E_{\rm min}$ relaxes, and one finds a smaller ratio $E_{\rm min}/E_{\rm cut}$, and therefore a spectrum spanning a larger energy range. In other words, if $M_+$ is too low for a fixed value of $E_{\rm cut}$, it is not possible to satisfy the constraint of the total power (Equation~[\ref{eq:power}]). To solve this problem, we decreased $\eta_{\rm p,max}$ systematically until we found a solution. Comparison of panel~(a) and panel~(b) indicates that a smaller value of $\eta_{\rm p,max}$ will relax the power constraint, so that solutions may be found for larger regions in $(M_+,E_{\rm cut})$ space.

With the solutions of source spectra in hand for the population of 24 BWs and RBs considered (Tables~\ref{tab1} and~\ref{tab2}), we may next calculate their transport through the Galaxy (Section~\ref{sec:transport}).

\subsection{Neglecting the primary component from population-synthesis MSPs}
\label{sec:prim}
We have noted that the secondary component almost always vastly dominates the primary component in the case of the BWs / RBs (Section~\ref{sec:BW2}), i.e., usually $M_+\gg 1$. This is due to the fact that multiplicities grow very rapidly with $\varepsilon$. Even in the case of $\varepsilon=0.0$, while the primary spectra may dominate the secondary spectra for some low-$B_{\rm s}$ and large-$P$ pulsars (which would imply $M_+\ll 1$), there will always be pulsars with high enough $B_{\rm s}$ and short $P$ (i.e., $M_+\sim100-1000$) so that their secondary spectra will dominate the cumulative flux contribution from a population of pulsars. This means that the cumulative spectrum from the BW and RB pulsars will be dominated by secondary, and not by primary spectra.\footnote{Neglecting the primary spectra in this case would imply setting $M_+ + 1 \approx M_+$ when solving Equation~(\ref{eq:current}). While we have not done this, the effect would be negligible, given the large values of $M_+$ in some cases, even for $\varepsilon=0.0$.}

On the other hand, for the MSPs from our population synthesis model, where we assume no shock acceleration, the primaries may form nearly mono-energetic spectra at very high Lorentz factors $\gamma\sim10^{7-8}$, depending on field-line curvature, i.e., colatitude, and also $P$ and $B_{\rm s}$. Given this small energy range (the spectrum is almost a $\delta$-distribution), one might think that this component may leave a distinct signature in the total spectrum of particles leaving the pulsar magnetosphere. However, when combining primary spectra from several pulsars, and following their transport through the Galaxy to Earth, the cumulative primary spectrum will have been smeared out due to the different source locations and properties. The primary spectra should also be at a lower intensity than the secondaries, given the typical multiplicities encountered for the $B_{\rm s}$ and $P$ values of the closest MSPs. Furthermore, if there would have been any signature at high energies $\sim10$~TeV, where the secondary spectra drop off in this case, this will be completely masked by the contribution of the BW / RB.

Given the above arguments, we do not include the primary spectra from the MSP synthesis population since they should not have an impact on our results. 


\section{Galactic transport of injected leptons}

\subsection{Interstellar Radiation Field}
\label{sec:ISRF}
Knowledge of the spectral and spatial properties of Galactic `background photons' is important for calculations of IC losses suffered by leptons propagating through our Galaxy. The relevant photons are optical ones produced by the population of stars in the Galaxy, in addition to infrared (IR) photons that are the result of scattering, absorption, and re-emission of the stellar photons by dust in the ISM; see e.g., \citet{Porter08}. 

The GALPROP code \citep{Strong98} includes a detailed model for this ISRF that incorporates a stellar population model (i.e., a luminosity distribution derived from 87 stellar / spectral classes distributed in 7 geometric locations within the Galaxy), dust grain abundance and size distribution models, as well as the absorption and scattering efficiencies of the latter which enable radiative transport calculations for stellar photons propagating through the ISM. While the ISRF is inherently anisotropic and inhomogeneous, with the bulk of the photons leaving the inner Galaxy, the ISRF model used by GALPROP assumes azimuthal symmetry and a cylindrical geometry. For more details, see \citet{Moskalenko06,Porter06,Porter08} and references therein. 

For our purposes, we only need average photon energy densities to calculate the total IC loss rates\footnote{However, the temperature $T_j$ of each blackbody component $j$ is needed when implementing KN corrections; see Equation~(\ref{eq:KN}).}, since this is the quantity needed to solve the transport equation (see Equation~[\ref{eq:transport}]). We find that the GALPROP ISRF is adequately approximated by three blackbody components \citep[optical, IR, and cosmic microwave background or CMB; see Figure~2 of][]{Venter15}. We follow \citet{Blies12} in distinguishing two main spatial regions: the Galactic Disk and the Galactic Halo. For the Disk, we use their values of $U_{\rm opt} = U_{\rm IR} = 0.4$~eV\,cm$^{-3}$, and $U_{\rm CMB} = 0.23$~eV\,cm$^{-3}$, which is similar to the values of \citet{Ruppel10}, while for the Halo, we use $U_{\rm opt} = 0.8$~eV\,cm$^{-3}$, $U_{\rm IR} = 0.05$~eV\,cm$^{-3}$, and $U_{\rm CMB} = 0.23$~eV\,cm$^{-3}$. The dust is assumed to follow the Galactic gas distribution \citep{Moskalenko06}, which tapers off strongly with perpendicular distance above the Galactic Plane, leading to less absorption of optical photons (and therefore a larger value for $U_{\rm opt}$ and a reduced value of $U_{\rm IR}$ in the Halo). 

\subsection{The Galactic Magnetic Field}
\label{sec:Bfield}
\citet{Han09} noted that there are five observational tracers of the Galactic magnetic field. These are polarization of starlight (indicating that the local field is parallel to the Galactic Plane and follows the local spiral arms); polarized thermal dust emission from molecular clouds (indicating field enhancement upon cloud formation via compression of the ISM, and that magnetic fields in these clouds seem to be preferentially parallel to the Galactic Plane); Zeeman splitting of spectral emission or absorption lines from molecular clouds or from OH masers associated with HII or star forming regions (indicating large-scale reversals in the sign of the line-of-sight component of the median field, and that interstellar magnetic fields are apparently preserved through the cloud and star formation processes); diffuse SR radio emission (used to estimate the total and ordered or regular field strength); and Faraday rotation of linearly polarized radiation from pulsars and extragalactic radio sources (giving a measure of strength and orientation of the line-of-sight component of the magnetic field). The combination of the latter with measurements of total intensity and the polarization vectors (from SR) allows one to distinguish between three field components: regular, anisotropic, and random \citep{Beck09}. 

For our purposes, we are interested in an average field strength that would determine the SR loss rate (Equation~[\ref{eq:SR}] below), and not so much in the overall Galactic field structure\footnote{\citet{Kistler12} raised the additional issue of particle transport in a turbulent magnetic field, which we will briefly consider in Section~\ref{sec:discuss}.} (which is still under debate). The total field has been estimated to be around 6~$\mu$G, averaged over a distance of 1~kpc around the Sun, using SR measurements and equipartition arguments where the magnetic energy density is set equal to that of cosmic rays. This number increases to $\sim10\,\mu$G closer to the inner Galaxy \citep[see][and references therein]{Beck09}. 
\citet{Han06} used a combination of dispersion and rotation measures of over 500 pulsars and found that the regular magnetic field decreases from $\sim6\,\mu$G near a Galactocentric distance of $2$~kpc to $\sim1\,\mu$G near $9$~kpc; the value is $\sim2\,\mu$G near the Sun \citep[see Figure~11 of][]{Han06}. 
The latter should be compared to recent measurements of the interstellar magnetic field by Voyager~2 which yielded $3.8-5.9\,\mu$G \citep{Burlaga14}. Furthermore, the mean regular field as function of latitude is inferred to vary between $\sim\pm5\,\mu$G \citep{Han06}. Fields in interarm regions are seemingly weaker than those in spiral arms. For example, the average regular field  in the Norma arm was found to be $4.4 \pm 0.9\,\mu$G \citep{Han02}. The regular magnetic field has only a weak vertical component of $B_z=0.2-0.3\,\mu$G, directed from the southern to the northern Galactic Pole \citep{HQ94}. 
\citet{Orlando13} inferred values of $\sim2\,\mu$G, $\sim5\,\mu$G, and $\sim2\,\mu$G for the local regular, random, and anisotropic field components in the Disk via Galactic SR modeling. The average total field, however, decreases when taking into account its rapid decay with height above the Plane. \citet[][hereafter D10]{D10} argue that SR losses depend on the mean of the squared magnetic field, so that one should include all components in the following way:
\begin{equation}
 B_{\rm SR} = \sqrt{\langle B_{\rm r}^2\rangle + \langle B_{\rm a}^2\rangle+\langle B_{\rm i}^2\rangle},
\end{equation}
with $B_{\rm r}$ the regular field, $B_{\rm a}$ the irregular, anisotropic field aligned with the regular one, and $B_{\rm i}$ the isotropic or random field; also, $\langle B_{\rm r}^2\rangle = \langle B_{\rm r}\rangle^2$. Results from \citet{Jaffe10} lead to values of up to $B_{\rm SR}\sim6\,\mu$G for fields in the Galactic Disk. However, if an exponential decay function for the vertical behaviour of the magnetic field is assumed, and $B_{\rm SR}$ is averaged over a spherical volume of radius 2~kpc, D10 finally obtains an average value of $B_{\rm SR}\sim1-3\,\mu$G.

\subsection{Total Leptonic Energy Loss Rate}
\label{sec:losses}
The SR loss rate is given by
  \begin{equation}
  \dot{E}_{\rm SR} = \frac{4\sigma_{\rm T}cU_BE^2}{3\left(m_ec^2\right)^2},\label{eq:SR}
  \end{equation}
with $m_e$ the electron mass, $E$ the particle energy, and $U_B$ the magnetic energy density
  \begin{equation}
  U_B = \frac{B^2}{8\pi} = 0.098b_2^2~{\rm eV\,cm}^{-3}
  \end{equation}
for a Galactic field of $b_2 = B/(2.0~\mu$G). The general expression (including KN effects) for the IC loss rate (for target photons of energy density $U_j$, and $j$ signifying different blackbody components associated with temperatures $T_j$) may be approximated as \citep[for details, see the Appendix of][]{Ruppel10}
  \begin{equation}
\dot{E}_{{\rm IC},j} = \frac{4\sigma_{\rm T}cU_jE^2}{3\left(m_ec^2\right)^2}\frac{\gamma_{{\rm KN},j}^2}{\gamma_{{\rm KN},j}^2 + \gamma^2},\label{eq:Ruppel}
  \end{equation}
with $\gamma$ the particle Lorentz factor, and the critical KN Lorentz factor defined as
  \begin{equation}
\gamma_{{\rm KN},j} \equiv \frac{3\sqrt{5}}{8\pi}\frac{m_ec^2}{k_BT_j} \approx \frac{0.27m_ec^2}{k_BT_j}.\label{eq:KN}
  \end{equation}
The IC loss rate for particles with Lorentz factors above $\gamma_{{\rm KN},j}$ is severely suppressed. If $\gamma\ll \gamma_{{\rm KN},j}$, we recover the well-known expression for the Thomson limit \citep{Blumenthal70}
  \begin{equation}
\dot{E}_{{\rm IC},j} = \frac{4\sigma_{\rm T}cU_jE^2}{3\left(m_ec^2\right)^2}.
\end{equation}
By considering various Galactic soft-photon target fields (IR, CMB, and optical) with respective energy densities $U_j$, we note that the KN correction is only necessary for optical photons, where $\gamma_{\rm KN,opt}\sim10^5$. 

Previously \citep{Venter15}, we assumed that we could approximate all losses as being in the Thomson limit for all cases. Since all loss terms (SR and IC, for the different soft-photon components) have the same functional dependence on energy in this case, $\dot{E}\propto E^2U$, where $U$ can indicate either magnetic or soft-photon energy density, we could define one single loss term using an effective magnetic field $B_{\rm eff}$ that takes into account both SR and IC losses. We found a value of $B_{\rm eff}\sim7\,\mu$G for both the Plane and the Halo, given the typical values used for $U_j$ and $B$. These $B_{\rm eff}$ values are the same in both regions because $U_{\rm opt}$ goes from 0.4~eV\,cm$^{-3}$ in the Plane to 0.8~eV\,cm$^{-3}$ in the Halo, while $U_{\rm IR}$ goes from 0.4~eV\,cm$^{-3}$ in the Plane to 0.05~eV\,cm$^{-3}$ in the Halo. In addition, the actual magnetic field drops from $B\sim3\,\mu$G to $B\sim1\,\mu$G \citep{Blies12}. In \citet{Venter15}, however, we decided to use a slightly lower value of $B_{\rm eff}=5\,\mu$G in view of the fact that the Thomson limit would overestimate the losses. 

For this paper, we introduce two loss terms, thereby separating those in the Thomson limit, and the one in the KN limit. We denote this as follows:
\begin{equation}
 \dot{E}_{\rm total} = \left(\dot{E}_{\rm Thom}\right) + \dot{E}_{\rm KN} = \left(\dot{E}_{\rm SR}+\dot{E}_{\rm IC,IR}+\dot{E}_{\rm IC,CMB}\right)  + \dot{E}_{\rm IC,opt}.
\end{equation}

To calculate $\dot{E}_{\rm Thom}$, we formally set $U_{\rm opt} = 0$~eVcm$^{-3}$. We can combine the rest of the terms into one by defining an effective magnetic field, $B_{\rm eff}$, as before:
  \begin{eqnarray}
  \dot{E}_{\rm Thom} &=& \dot{E}_{\rm SR} + \dot{E}_{\rm IC,IR} + \dot{E}_{\rm IC,CMB} = \frac{4\sigma_{\rm T}cE^2}{3\left(m_ec^2\right)^2}\left[U_B + U_{\rm IR} + U_{\rm CMB} \right]\nonumber\\
  & = &\frac{4\sigma_{\rm T}cE^2}{3\left(m_ec^2\right)^2}U_{\rm eff} = b_0E^2,\\
  b_0 & = & \frac{4c}{9}\left(\frac{e}{m_ec^2}\right)^4B_{\rm eff}^2 = 1.58\times10^{-15}\left(\frac{B_{\rm eff}}{1~\mu{\rm G}}\right)^2.
  \end{eqnarray}
We find values of $B_{\rm eff} = 5.2 - 5.9\,\mu$G in the Plane, and $B_{\rm eff} = 3.6 - 4.6\,\mu$G in the Halo (the range stemming from the fact that we considered $B$ to be in the range $1-3\mu$G; see Section~\ref{sec:Bfield}).

We have to treat the optical component separately, since the KN effect will become important in this case. Following \citet{Blies12}, we consider only the most dominant optical component, which we approximate by a black body with $T_{\rm opt} = 5,000$~K, replacing $U_j$ by $U_{\rm opt}$ and $\gamma_{{\rm KN},j}$ by $\gamma_{\rm KN,opt}$ in Equation~(\ref{eq:Ruppel}):
  \begin{equation}
\dot{E}_{\rm KN} = \frac{4\sigma_{\rm T}cU_{\rm opt}E^2}{3\left(m_ec^2\right)^2}\frac{\gamma_{\rm KN,opt}^2}{\gamma_{\rm KN,opt}^2 + \gamma^2}.
  \end{equation}
The particles will traverse regions having different (line-of-sight-averaged) $U_{\rm opt}$. Furthermore, the optical stellar model of \citet{Wainscoat92} used to calculate the ISRF gives a scale height of $0.27- 0.325$~kpc for the stars of $T_{\rm opt}\sim5,000$~K, while the scale height for the Galactic magnetic field varies between $\sim0.1 - 4$~kpc \citep[D10; ][]{Orlando13}. In view of the uncertainties associated with obtaining a line-of-sight-averaged $B_{\rm eff}$ and $U_{\rm opt}$ for each source, and given the fact that our transport model considers only one spatial dimension, in what follows we consider two extreme cases of minimal and maximal total losses $\dot{E}_{\rm total}$ to bracket our particle flux results: (1) $B_{\rm eff} = 3.6\,\mu$G and $U_{\rm opt} = 0.4$~eV\,cm$^{-3}$; and (2) $B_{\rm eff} = 5.9\,\mu$G and $U_{\rm opt}$ = 0.8~eV\,cm$^{-3}$.

\subsection{Solution of the transport equation}
\label{sec:transport}
In order to transport the pairs created in the MSP magnetospheres to Earth, we use the following Fokker-Planck-type equation that includes spatial diffusion and energy losses:
  \begin{equation}
\frac{\partial n_{\rm e}}{\partial t}=\mathbf{\nabla}\cdot\left({\cal K}\cdot\mathbf{\nabla} n_{\rm e} \right)-\frac{\partial}{\partial E}\left(\dot E_{\rm total} n_{\rm e}\right)+S,\label{eq:transport}
  \end{equation}
with $n_{\rm e}$ the lepton density (per energy interval). Also, ${\cal K}$ denotes the diffusion tensor and $\dot E_{\rm total}$ the total energy losses, while $S$ is the source term. 

Since MSPs are quite old (ages of $\sim10^{10}$~yr), and have very small time derivatives of their period $\dot{P}$, we assume a steady-state scenario ($\partial/\partial t=0$). We furthermore assume a uniform ISM, and thus invoke spherical symmetry such that $n_{\rm e}$ only depends on distance $r=d$ from Earth. For a scalar diffusion coefficient $\kappa$, Equation~(\ref{eq:transport}) now reduces to
\begin{equation}
0=\frac{1}{{r}^2}\frac{\partial}{\partial r}\left({r}^2 \kappa \frac{\partial n_{\rm e}}{\partial r}\right) -\frac{\partial}{\partial E}\left(\dot E_{\rm total } n_{\rm e}\right)+Q.
\end{equation}

We incorporate the energy losses as explained in Section~\ref{sec:losses} and assume that the diffusion coefficient is spatially independent (so that ${\cal K}$ becomes a function of energy only, $\kappa(E)$) and we assume a power law energy dependence and (as motivated by quasi-linear theory; see, e.g., \citealt{Maurin02})
  \begin{equation}
  \kappa(E) = \kappa_0\left(\frac{E}{E_{\rm norm}}\right)^{\alpha_{\rm D}}.\label{eq:alpha_D}
  \end{equation}
We assume typical values of $\alpha_{\rm D} = 0.6$, $E_{\rm norm} = 1$~GeV, and $\kappa_0 = 0.1~{\rm kpc}^2{\rm Myr}^{-1} \approx 3\times10^{28}$~cm$^2$s$^{-1}$ \citep[e.g.,][]{Moskalenko98,Malyshev09,Grasso09,Feng15}. The value for $\kappa_0$ is an indication of the efficiency of the diffusion process at a particular energy \citep[e.g.,][]{Maurin02}, while $\alpha_{\rm D}$ is inferred from the measured ratio of boron to carbon and characterizes the escape time of cosmic rays from the Galaxy \citep[e.g.,][]{Blasi09,Genolini15}.

For the source term, we consider $N\sim5\times10^4$ Galactic MSPs from the population synthesis code (Section~\ref{sec:synth}), and $N=24$ for the BW / RB case (Section~{\ref{sec:BW}}). For the $i^{\rm th}$ pulsar in our synthesis population, we assign a pair spectrum $Q_i(P,B_{\rm s},\varepsilon,E)$, as calculated in Section~\ref{sec:pairs} for the corresponding simulated values of $P$, $B_{\rm s}$, and $\varepsilon$. We model this as
\begin{equation}
    S = \sum_i^NQ_i(P,B_{\rm s},E)\delta(\mathbf{r} - \mathbf{r}_{0,i}).
  \end{equation}
Here, $\mathbf{r}_{0,i}$ are the source positions. For an infinite system, Equation~(\ref{eq:transport}) is solved by the following Green's function \citep[e.g., D10;][]{Blies12}:
\begin{equation}
G(\mathbf{r},\mathbf{r}_0,E,E_0) = \frac{\Theta(E_0 - E)}{\dot{E}_{\rm total}\left(\pi\lambda\right)^{3/2}}\exp\left(-\frac{|\mathbf{r} - \mathbf{r}_0|^2}{\lambda}\right), \label{eq:Greens}
\end{equation}
with $E_0$ the particle energy at the source, and the square of the propagation scale is characterized by 
\begin{eqnarray}
\lambda(E,E_0) & \equiv & 4\int_{E}^{E_0}\frac{\kappa(E^\prime)}{\dot E_{\rm total}(E^\prime)}\,dE^\prime,\\
 & = & \lambda_0\left[\frac{1}{E_0}\left(\frac{E_0}{E_{\rm norm}}\right)^{\alpha_{\rm D}} - \frac{1}{E}\left(\frac{E}{E_{\rm norm}}\right)^{\alpha_{\rm D}}\right],\\
\lambda_0 & = & \frac{4\kappa_0}{\left(\alpha_{\rm D}-1\right)b_0},
\end{eqnarray}
and $\Theta(E_0 - E) $ the Heaviside function. The latter is used to ensure that $\lambda>0$.
The lepton flux may then be found using
\begin{equation}
\phi_e(\mathbf{r},E) = \frac{c}{4\pi}\int\!\!\!\!\int\!\!\!\!\int\!\!\!\!\int G(\mathbf{r},\mathbf{r}_0,E,E_0)S\,dE_0d^3r_0.\label{eq:int}
\end{equation}
While the finite boundary of the Galactic Halo should impact the solution, this effect is not too large for GeV leptons, for which the propagation scale is only a few kpc (D10), and we neglect it here for simplicity. Our results will indicate that our predicted MSP contribution becomes significant above $\sim10$~GeV, so that the effect of solar modulation may safely be neglected \citep{Strauss14}. Indeed, \citet{Accardo14} noted that modulation has no effect on the newly measured PF by \textit{AMS$-$02}, although \citet{Aguilar14} claimed that they see the effects of solar modulation up to $\sim10$~GeV in their electron and positron data.

We found that in order to have smooth output spectra, we had to treat nearby and distant sources separately. We used a logarithmic grid for the particle source energies $E_0$ of the distant sources ($d>1$~kpc), which is strongly refined in $E_0$ for the nearby ($d<1$~kpc) ones ($\sim 100$ sources) when solving Equation~(\ref{eq:int}). This was necessary, since there are ``poles'' in the Green's function when $E\approx E_0$, and $\lambda\approx 0$, so $G(\mathbf{r},\mathbf{r}_0,E,E_0) \rightarrow \infty$ (Equation~[\ref{eq:Greens}]). These singularities are however, removable, in the sense that a very fine grid in $E_0$ results in a finite integrand for Equation~(\ref{eq:int}), while the Heaviside function formally avoids $E_0 = E$.

\begin{figure}
\epsscale{1.0}
\plotone{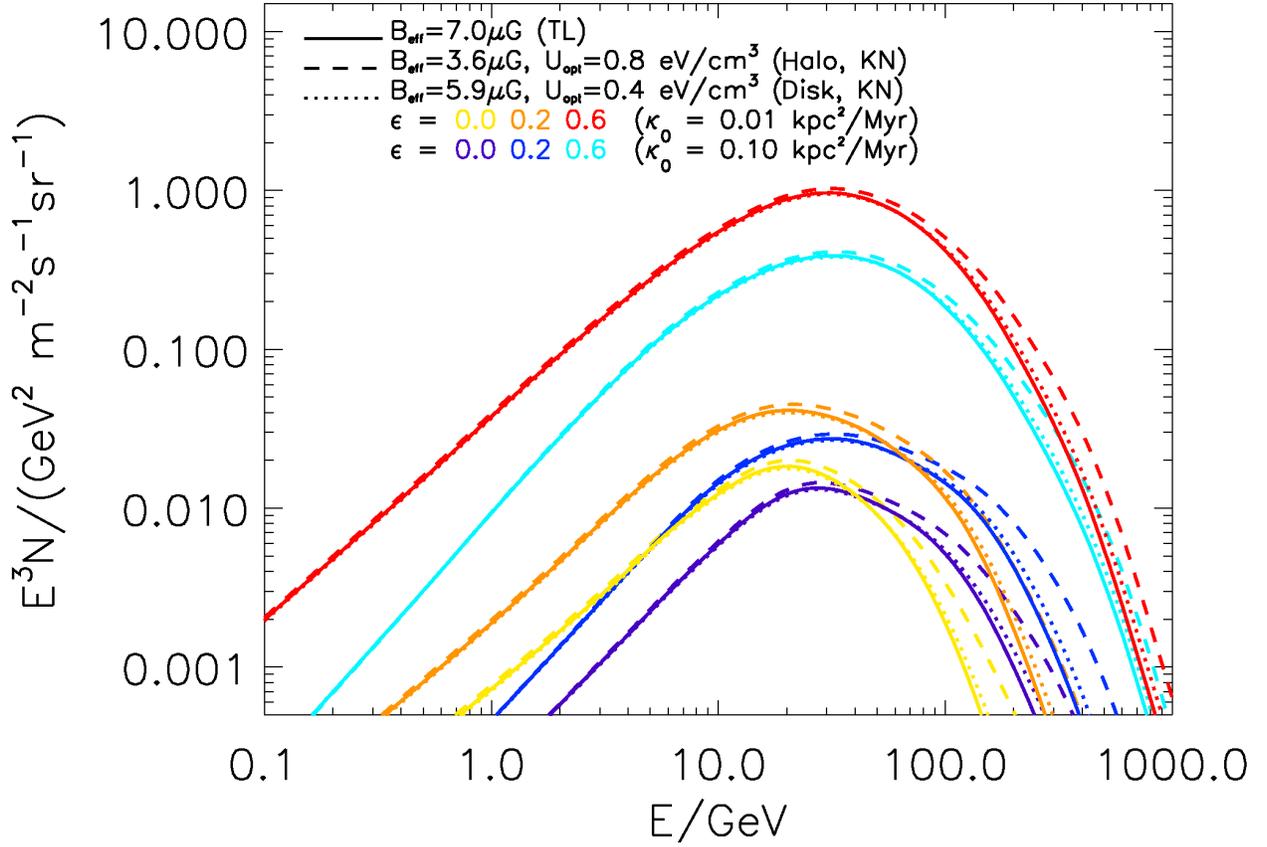}
\caption{Comparison of transport done for the Galactic synthesis component (for one particular realization of the MSP population using the synthesis code) in the Thomson (TL) and KN limits. We compare Disk and Halo scenarios (which have the same value of $B_{\rm eff} = 7\,\mu$G in the case of the Thomson limit), and also consider results for different values of $\varepsilon$ and $\kappa_0$, as indicated in the legend.}\label{fig:KN}
\end{figure}

Figure~\ref{fig:KN} shows a comparison of transported spectra involving the Galactic synthesis MSP component, in the Thomson limit (including all background photons, plus SR losses) vs.\ the KN limit (i.e., SR, Thomson limit for IR and CMB, but KN limit for the optical photons). See Section~\ref{sec:losses} for details. We compare Disk and Halo scenarios. In the Thomson limit, these imply the same value of $B_{\rm eff} = 7\,\mu$G (indicated by solid lines; different values for $\varepsilon$ are distinguished by the colors); however, in the KN limit, we have to separate the optical photon component, and we indicate the relevant values in the legend (dashed lines are for the Halo, for $B_{\rm eff} = 3.6\,\mu$G and $U_{\rm opt}$ = 0.8 eV\,cm$^{-3}$, while dotted lines indicate $B_{\rm eff} = 5.9\,\mu$G and $U_{\rm opt}$ = 0.4 eV\,cm$^{-3}$, for the Disk). It is noticable that the particles at higher energies suffer fewer losses in the KN case, given the reduction of the losses above $\sim160$~GeV in this regime, and hence this raises the transported spectrum somewhat. The largest enhancement of particle flux occurs for the Halo case, given the low value of $B_{\rm eff}$. We also show the effect of changing the normalization of the diffusion coefficient. For a larger $\kappa_0$ (cool colors), the flux is lower, while the opposite occurs for a smaller value of $\kappa_0$ (warm colors). One may view the latter case as a pile-up of particles, and one can also observe a transfer of high-energy particles to lower energies, given the slower diffusion, as evidenced by the change in slope at lower energies. Lastly, an increase in flux with $\varepsilon$ is evident, given the larger value of $M_+$ implied by a larger value of $\varepsilon$.

\section{Results}
\label{sec:results}
\begin{figure}
\epsscale{1.0}
\plotone{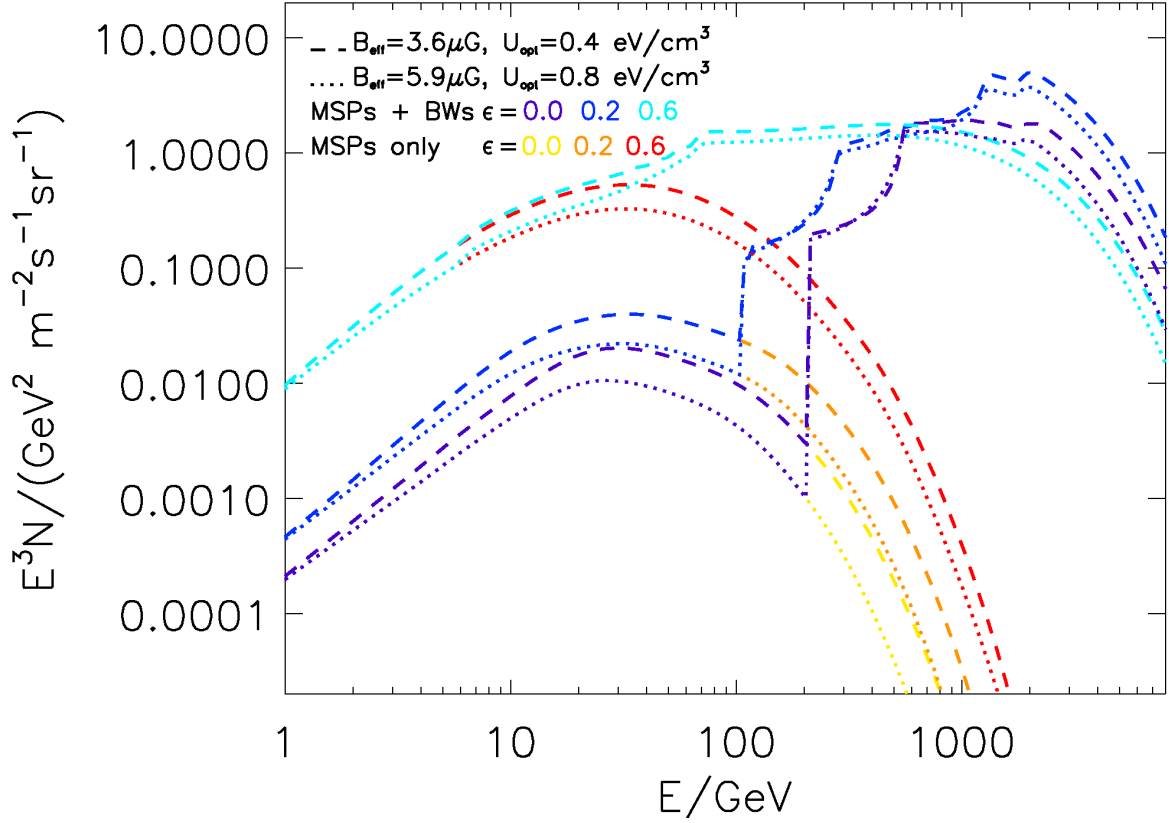}
\caption{Comparison of contribution of synthesis vs.\ BW / RB component, assuming $\kappa_0 = 0.1$~kpc$^2$\,Myr$^{-1}$ and $\eta_{\rm p,max}=0.1$.}\label{fig:fluxes2a}
\end{figure}

\begin{figure}
\epsscale{1.0}
\plotone{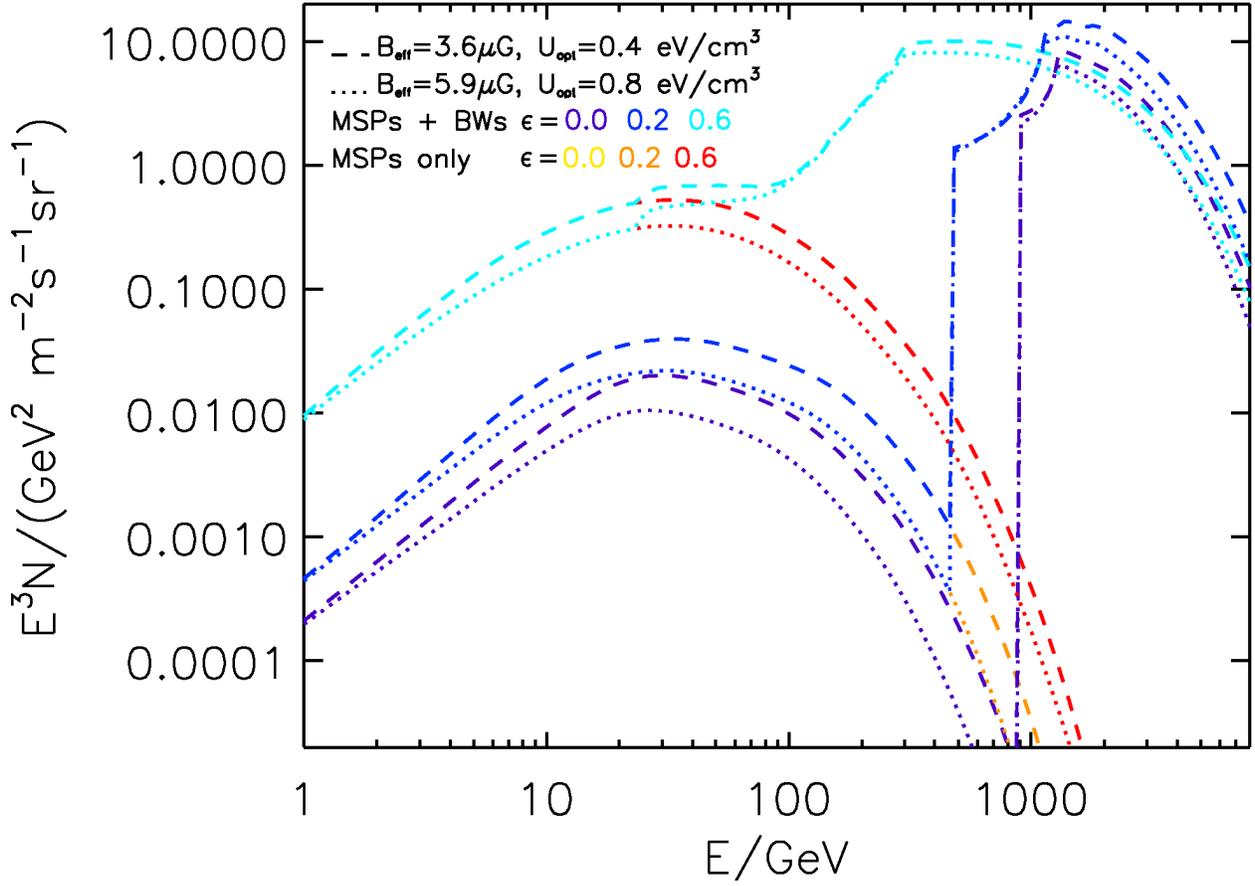}
\caption{Same as Figure~\ref{fig:fluxes2a}, but assuming $\kappa_0 = 0.1$~kpc$^2$\,Myr$^{-1}$ and $\eta_{\rm p,max}=0.3$.}\label{fig:fluxes2b}
\end{figure}

Figure~\ref{fig:fluxes2a} shows the synthesis and BW / RB spectra transported to Earth, as well as the sum of the synthesis and BW / RB components (see legend). Dashed lines are for $B_{\rm eff} = 3.6\,\mu$G and $U_{\rm opt}$ = 0.4 eV\,cm$^{-3}$, while dotted lines indicate $B_{\rm eff} = 5.9\,\mu$G and $U_{\rm opt}$ = 0.8 eV\,cm$^{-3}$ (i.e., minimal and maximal particle losses). The different values for $\varepsilon$ are indicated by different colors as noted in the legend, and we assumed $\kappa_0 = 0.1$~kpc$^2$\,Myr$^{-1}$ and $\eta_{\rm p,max}=0.1$. Figure~\ref{fig:fluxes2b} is the same, but for $\eta_{\rm p,max}=0.3$. While the synthesis component contributes mostly at tens of GeV, the BW / RB component contributes at thousands of GeV. It is also noticable that the BW / RB contribution is higher in this case due to the larger maximum shock efficiency. The discontinuous jump in the total spectrum is caused by the fact that the BW / RB component is the sum of a small number of spectra that cut off at particular values of the minimum particle energy $E_{\rm min}$ (calculated in each case by suitable normalization of the various binary injection spectra; see Section~\ref{sec:BW2}), and that the flux of this component dominates over that of the synthesis component (which cuts off around $\sim30$~GeV), making these low-energy cutoffs more evident. Furthermore, the spectral variations at low energies for the BW / RB component may be attributed to the fact that we are adding only 24 detected sources to obtain (a lower limit of) the cumulative contribution of binary MSPs to the cosmic-ray flux. Such variations should be smoothed out if a larger number of sources is used in this calculation.

\begin{figure}
\epsscale{1.0}
\plotone{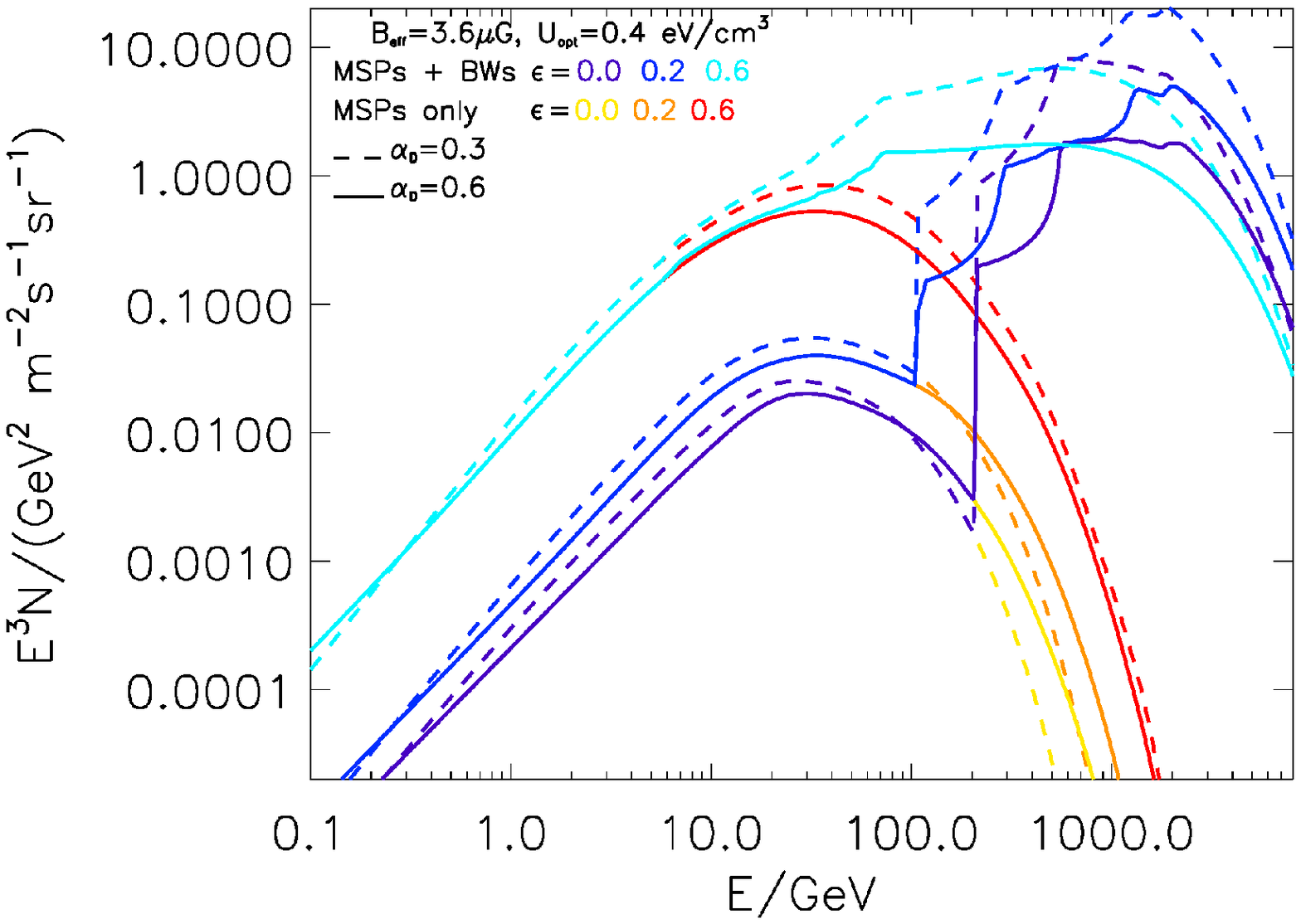}
\caption{Same as Figure~\ref{fig:fluxes2a}, but for $B_{\rm eff} = 3.6\,\mu$G, $U_{\rm opt}$ = 0.4 eV\,cm$^{-3}$, $\eta_{\rm p,max} = 0.1$,  $\kappa_0 = 0.1$~kpc$^2$\,Myr$^{-1}$, and $\alpha_{\rm D} = 0.3$ and $0.6$.}\label{fig:alpha_D}
\end{figure}

We next investigated the effect of varying the energy dependence of the diffusion coefficient by varying the parameter $\alpha_{\rm D}$ (see Equation~[\ref{eq:alpha_D}] and Figure~\ref{fig:alpha_D}). We fixed $B_{\rm eff} = 3.6\,\mu$G, $U_{\rm opt}$ = 0.4 eV\,cm$^{-3}$, also setting $\eta_{\rm p,max} = 0.1$, $\kappa_0 = 0.1$~kpc$^2$\,Myr$^{-1}$, and choosing values of $\alpha_{\rm D} = 0.3$ and 0.6, given the uncertainty of this parameter. (In the rest of the paper, we fix $\alpha_{\rm D} = 0.6$, unless stated otherwise.) Different values of $\varepsilon$ are indicated in the legend, as in previous plots. A smaller value of $\alpha_{\rm D}$ corresponds to relatively lower diffusion coefficients above the break energy of $E_{\rm norm} = $1~GeV. This has the same effect as assuming a smaller normalization $\kappa_0$, i.e., the spectra at Earth are relatively higher due to increased particle density. This effect is even more evident at higher energies, where the BW / RB component dominates, leading to significant uncertainties in this component's flux. The situation is opposite for source particle energies smaller than $E_{\rm norm}$, and one can see the opposite effect at terrestrial particle energies below $\sim200$~MeV.

\begin{figure}
\epsscale{1.0}
\plotone{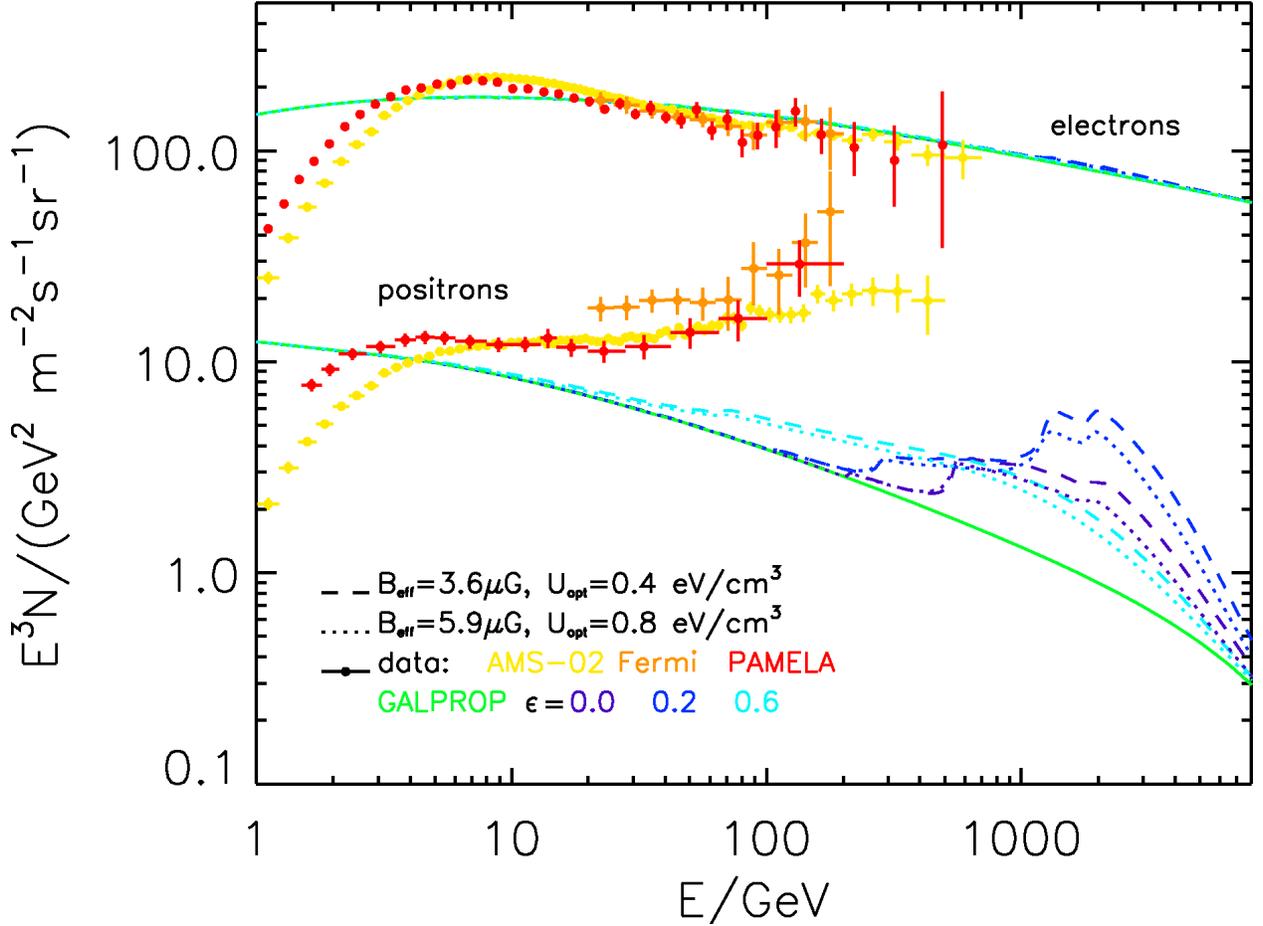}
\caption{Total MSP contribution (assumed to be equal numbers of positrons and electrons) to the leptonic cosmic-ray spectrum at Earth, assuming $\kappa_0 = 0.1$~kpc$^2$\,Myr$^{-1}$ and $\eta_{\rm p,max}=0.1$. Electron spectra appear at the top, while positron spectra appear lower down. The contribution from the synthesis component to the positron spectrum is visible at $\sim30$~GeV (for $\varepsilon=0.6$), and that of the BWs and RBs at $\sim1$~TeV. The cool colors (purple, blue, and cyan) indicate spectra for $\varepsilon = 0.0, 0.2,$ and $0.6$. Green indicates the ``background'' (non-MSP) electrons and positrons using output from the GALPROP code \citep{Vladimirov11} for standard parameters. Also shown are data from \textit{PAMELA} \citep[red;][]{Adriani13}, \textit{Fermi} \citep[orange;][]{Ackermann12}, and \textit{AMS$-$02} \citep[yellow;][]{Aguilar14}, accessed via the website http://lpsc.in2p3.fr/cosmic-rays-db \citep{Maurin14}.}\label{fig:LIS1}
\end{figure}

Figure~\ref{fig:LIS1} indicates the ``background'' electron and positron fluxes predicted by GALPROP\footnote{http://galprop.stanford.edu/webrun/} \citep{Vladimirov11} for standard parameters, as well as data from \textit{Fermi} \citep{Ackermann12}, \textit{PAMELA} \citep{Adriani13}, and \textit{AMS$-$02} \citep{Aguilar14}. We indicate synthesis spectra plus BW / RB spectra (we assume equal numbers of positrons and electrons) for dipole offsets of $\varepsilon = (0.0, 0.2, 0.6)$ and combinations of $(B_{\rm eff},U_{\rm opt}) = (3.6\,\mu$G,\,0.4 eV\,cm$^{-3}$) and $(B_{\rm eff},U_{\rm opt}) = (5.9\,\mu$G,\,0.8 eV\,cm$^{-3}$), i.e., minimal and maximal losses. The various curves are distinguished in the Figure caption. We set $\kappa_0 = 0.1$~kpc$^2$\,Myr$^{-1}$, $\eta_{\rm p,max}=0.1$. 

\begin{figure}
\epsscale{1.0}
\plotone{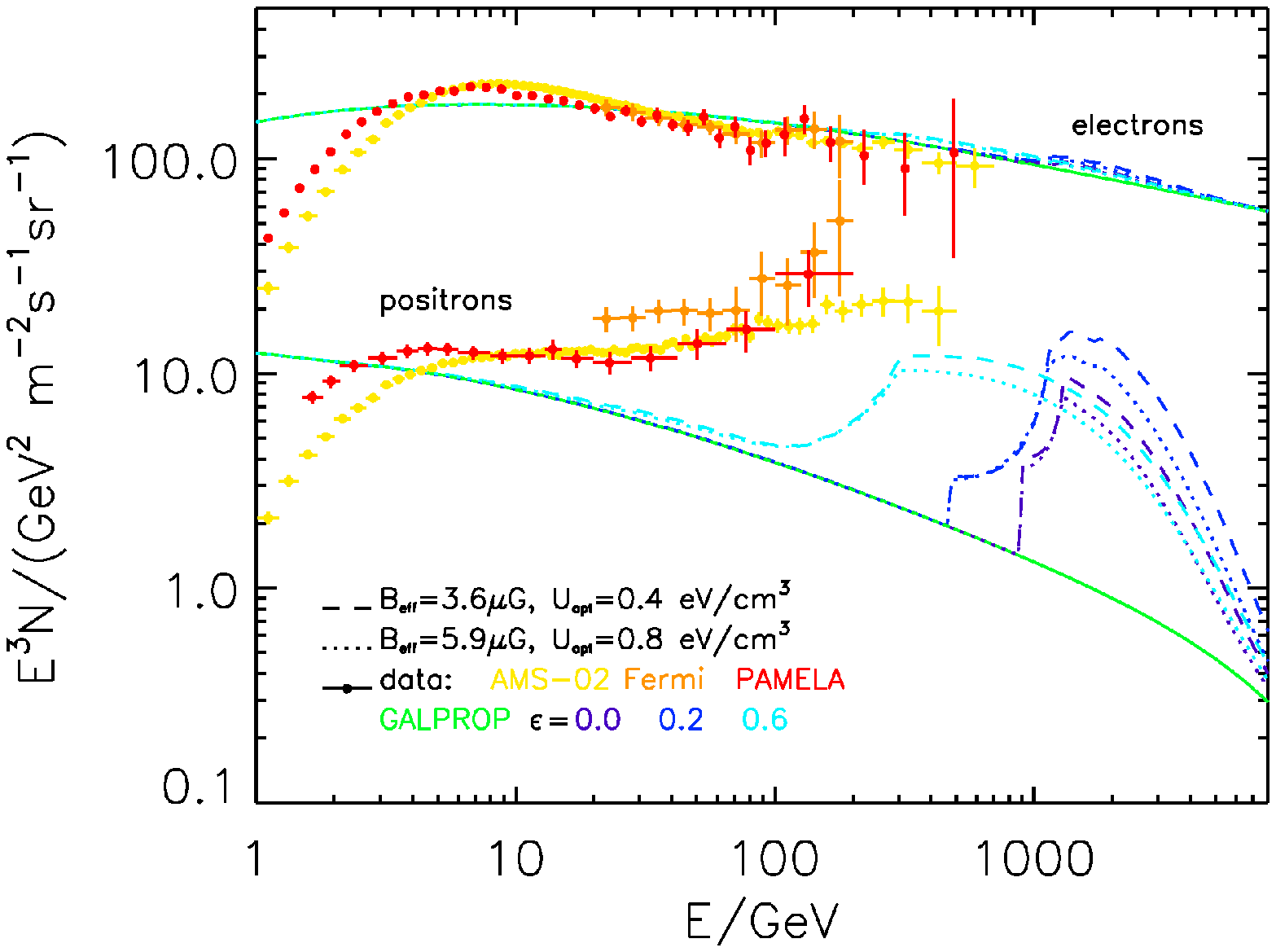}
\caption{Same as Figure~\ref{fig:LIS1}, but for $\kappa_0 = 0.1$~kpc$^2$\,Myr$^{-1}$ and $\eta_{\rm p,max}=0.3$.}\label{fig:LIS2}
\end{figure}

\begin{figure}
\epsscale{1.0}
\plotone{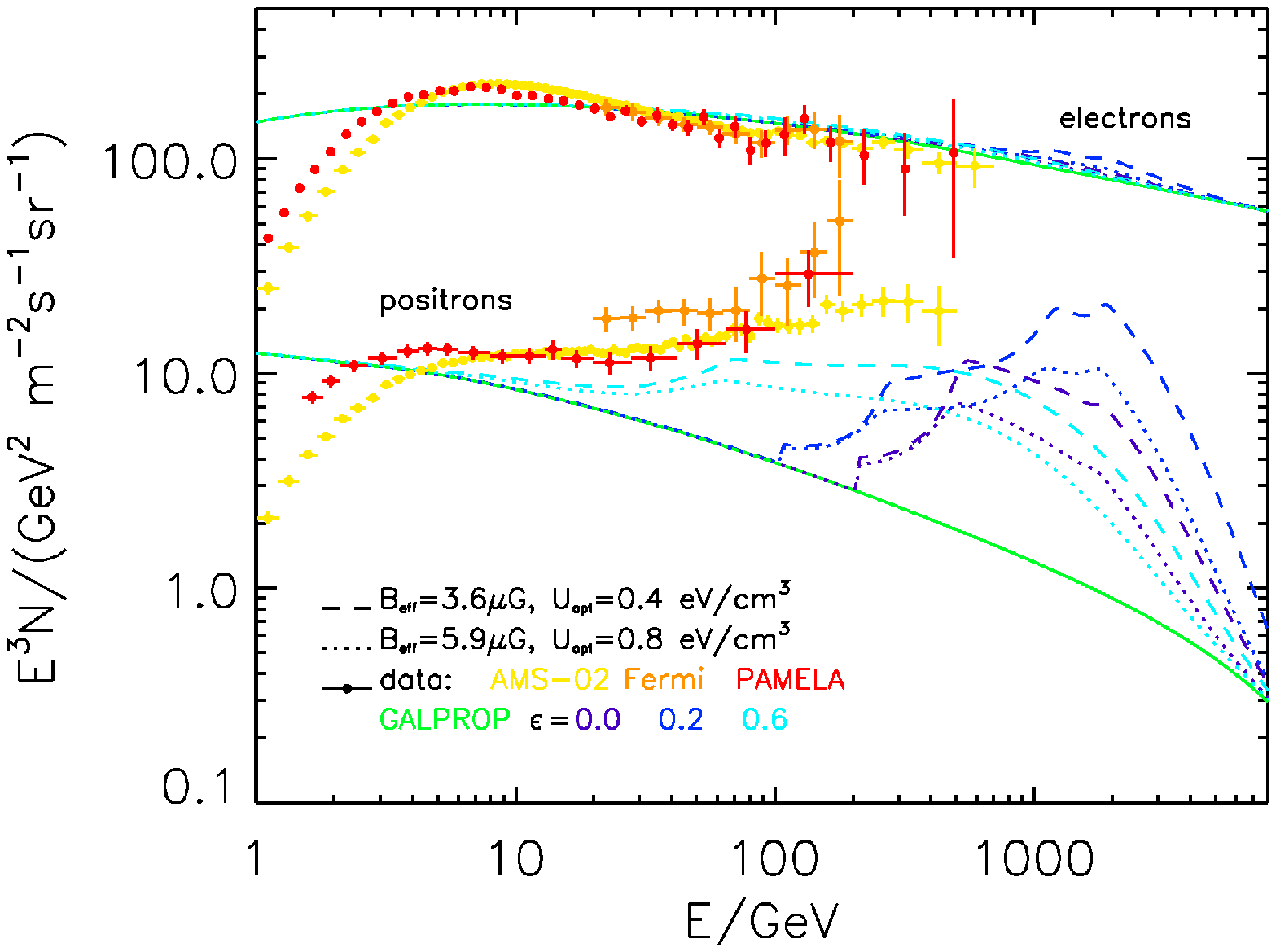}
\caption{Same as Figure~\ref{fig:LIS1}, but for $\kappa_0 = 0.01$~kpc$^2$\,Myr$^{-1}$ and $\eta_{\rm p,max}=0.1$.}\label{fig:LIS3}
\end{figure}

\begin{figure}
\epsscale{1.0}
\plotone{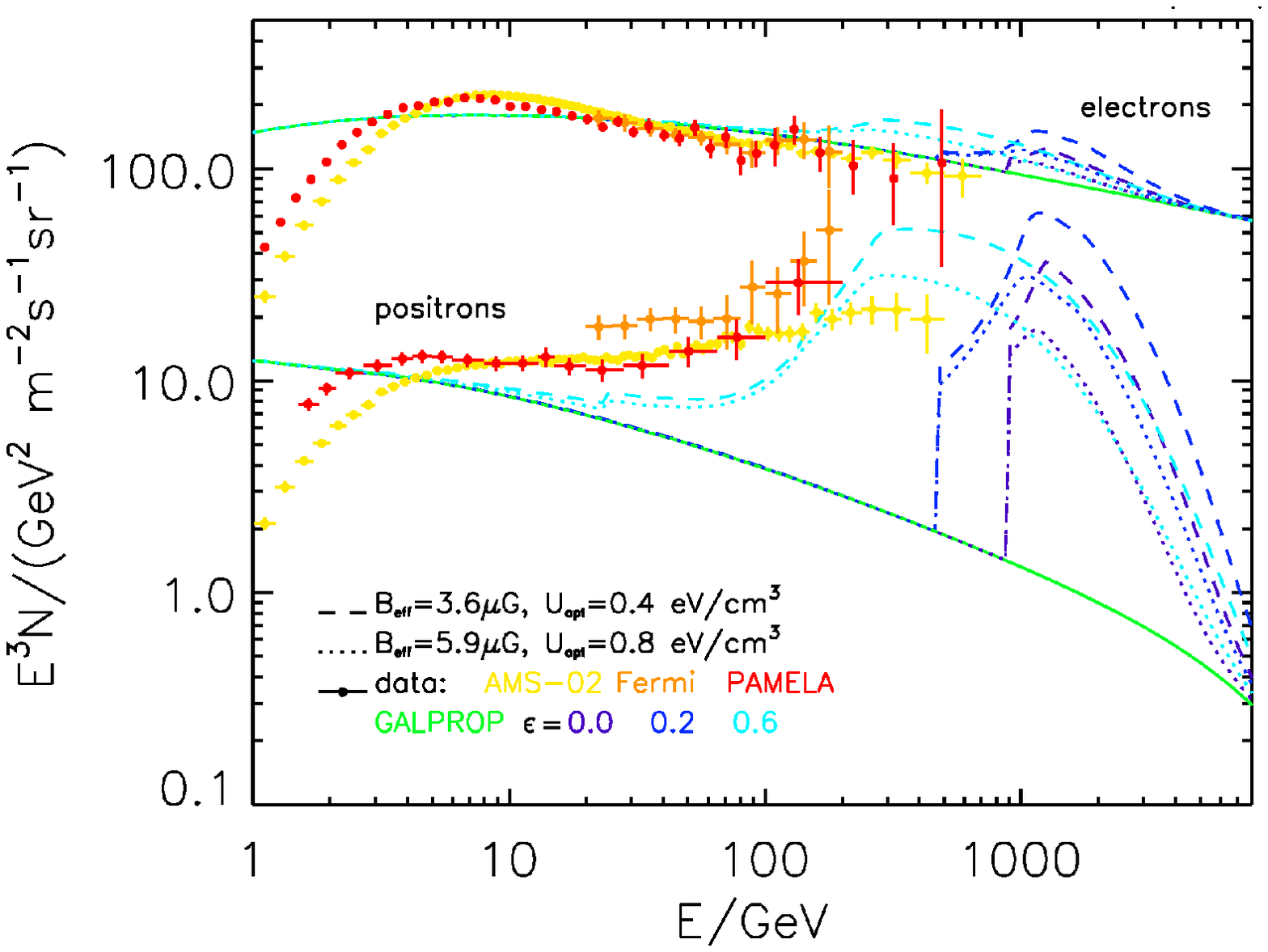}
\caption{Same as Figure~\ref{fig:LIS1}, but for $\kappa_0 = 0.01$~kpc$^2$\,Myr$^{-1}$ and $\eta_{\rm p,max}=0.3$.}\label{fig:LIS4}
\end{figure}

\begin{figure}
\epsscale{1.0}
\plotone{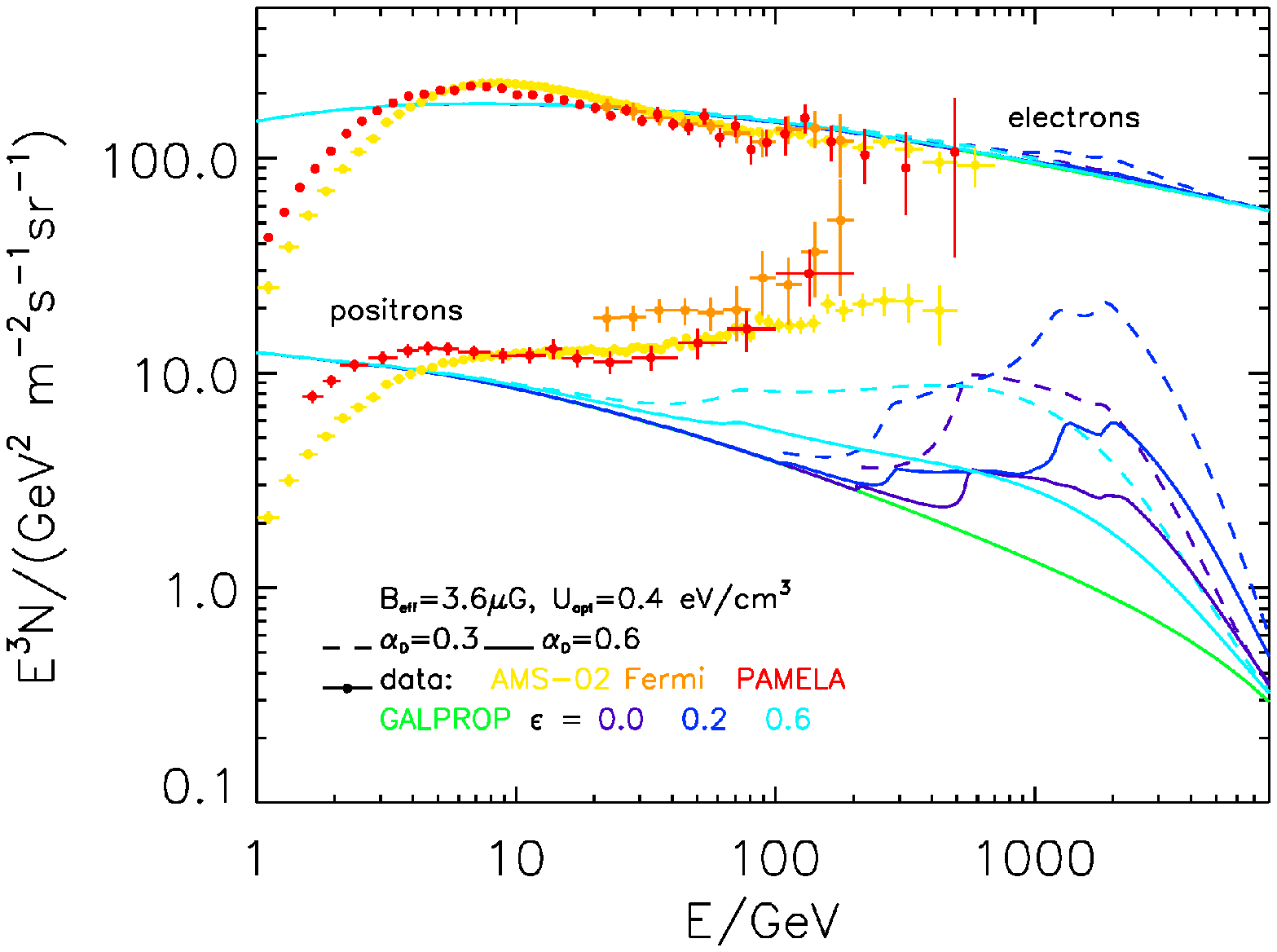}
\caption{Same as Figure~\ref{fig:LIS1}, but for $B_{\rm eff} = 3.6\,\mu$G, $U_{\rm opt}$ = 0.4 eV\,cm$^{-3}$, $\eta_{\rm p,max} = 0.1$,  $\kappa_0 = 0.1$~kpc$^2$\,Myr$^{-1}$, and $\alpha_{\rm D} = 0.3$ and $0.6$.}\label{fig:LIS4a}
\end{figure}

\begin{figure}
\epsscale{1.0}
\plotone{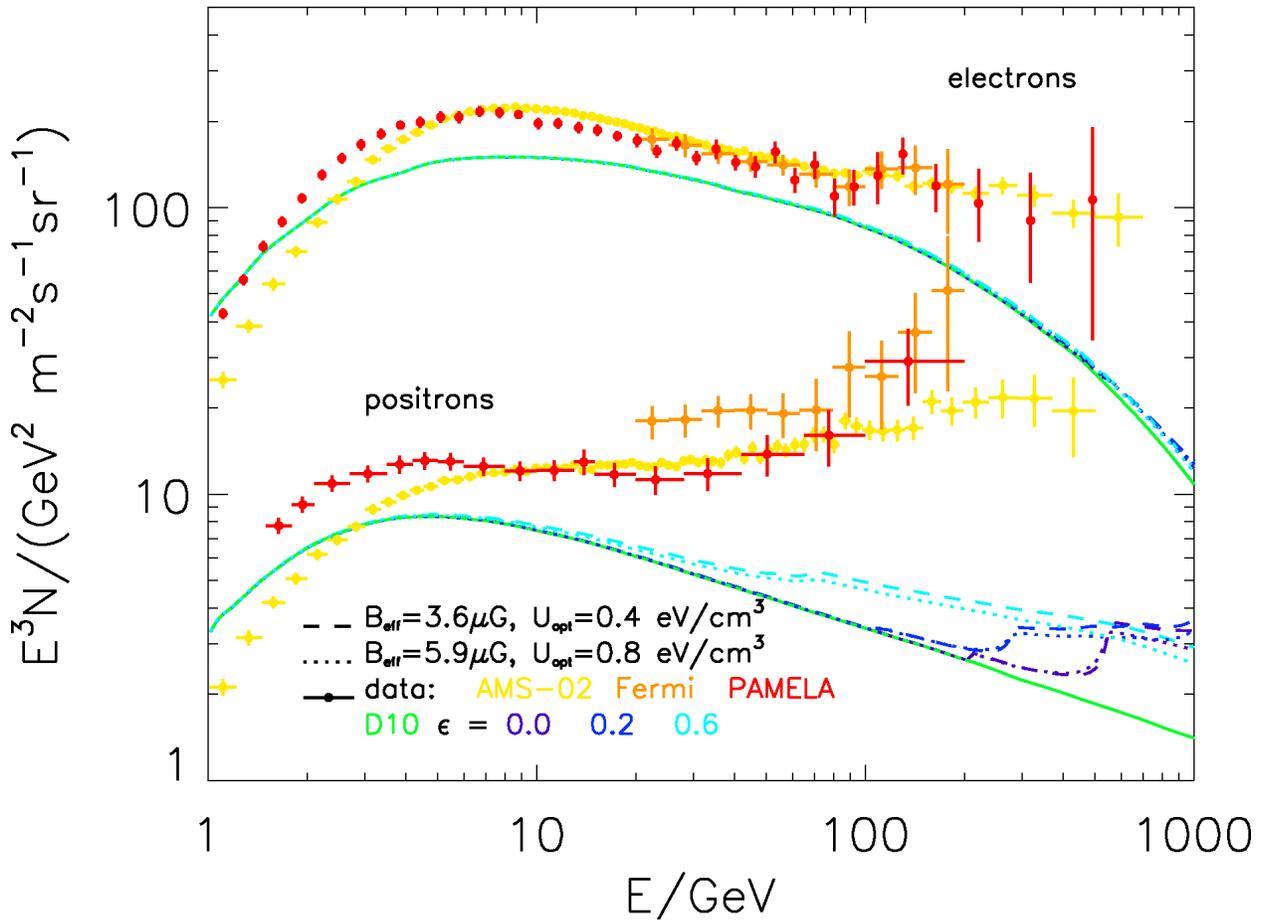}
\caption{Same as Figure~\ref{fig:LIS1}, but for the background model of D10.}\label{fig:LIS5}
\end{figure}

Figure~\ref{fig:LIS2} indicates the case for $\kappa_0 = 0.1$~kpc$^2$\,Myr$^{-1}$, $\eta_{\rm p,max}=0.3$, while Figure~\ref{fig:LIS3} and Figure~\ref{fig:LIS4} are for $\kappa_0 = 0.01$~kpc$^2$\,Myr$^{-1}$, $\eta_{\rm p,max}=0.1$ and $\kappa_0 = 0.01$~kpc$^2$\,Myr$^{-1}$, $\eta_{\rm p,max}=0.3$, respectively. As before, we see that the BW / RB contribution is higher for a larger shock efficiency, and that all components are higher for a smaller diffusion coefficient. This is due to a pile-up effect which boosts the particle density. We indicate the effect of changing $\alpha_{\rm D}$ in Figure~\ref{fig:LIS4a}, where one can see that the flux increases for smaller values of $\alpha_{\rm D}$, as noted earlier. For comparison, we show in Figure~\ref{fig:LIS5} the results when using the background model of D10 (we use their secondary positron flux as well as the sum of their secondary electron flux and primary electron flux originating in distant SNRs, as indicated in their Figure~14), where we assume $\kappa_0 = 0.1$~kpc$^2$\,Myr$^{-1}$, $\eta_{\rm p,max}=0.1$. The shape of the background model can strongly influence the total lepton spectrum.

For completeness, we wanted to test the synthesis model prediction against that obtained using detected radio pulsars, to ensure that the first is indeed higher, since it encapsulates both detected and undetected pulsars. We selected all pulsars with $P<0.1$~s, $\dot{P}>0$, and $d<2$~kpc from the ATNF Pulsar Catalog\footnote{http://www.atnf.csiro.au/people/pulsar/psrcat/} \citep{ATNF05}. We removed globular cluster pulsars, young pulsars (such as Vela), and known BW and RB systems by hand, leaving us with $\sim80$ MSPs. We used Shklovskii-corrected values for $\dot{P}$ \citep{Shklovskii70} when available. We then repeated the calculation above, and plotted the result (not shown) in order to compare with that from the population synthesis (where we have $\sim100$ sources within 1~kpc). We confirmed that the ``ATNF component'' was lower than the ``synthesis component'', as expected, since the detected pulsars should be a lower limit to the total number of sources predicted by the synthesis model. However, this is strongly dependent on $\varepsilon$, with the ``ATNF component'' becoming closer to the ``synthesis component'' for higher $\varepsilon$. This reflects the facts that the main contribution comes from nearby, powerful MSPs, and that the dominant contribution comes from pairs, the level of which very sensitively depends on pair multiplicity.

\begin{figure}
\epsscale{1.0}
\plotone{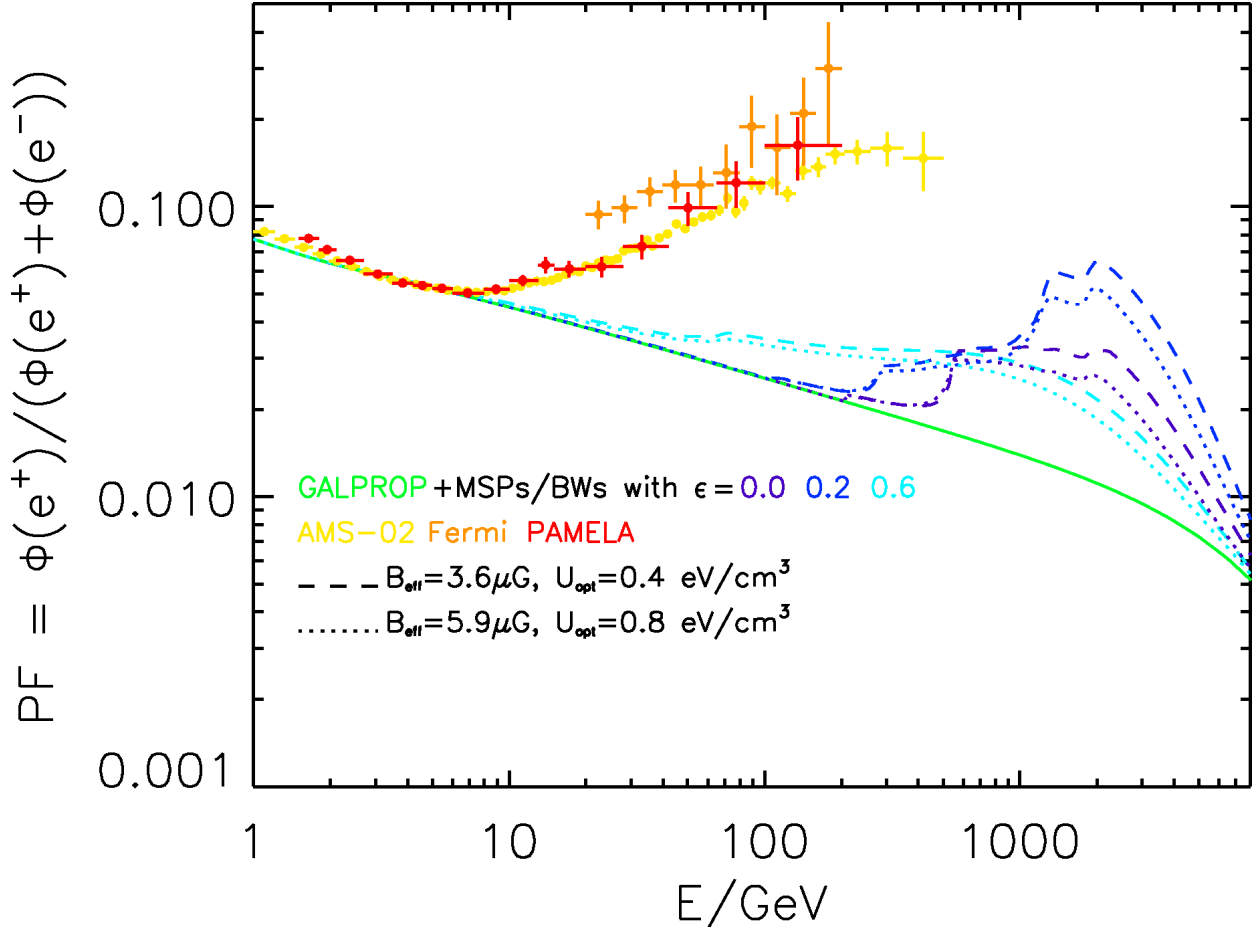}
\caption{Measured \citep{Ackermann12,Adriani13,Accardo14} and predicted PF (including ``background'' contributions from GALPROP in green and the synthesis plus BW / RB contributions from this work in purple, blue, and cyan, indicating $\varepsilon = (0.0, 0.2,$ and $0.6$). Here, $\kappa_0 = 0.1$~kpc$^2$\,Myr$^{-1}$ and $\eta_{\rm p,max} = 0.1.$}\label{fig:PF1}
\end{figure}

\begin{figure}
\epsscale{1.0}
\plotone{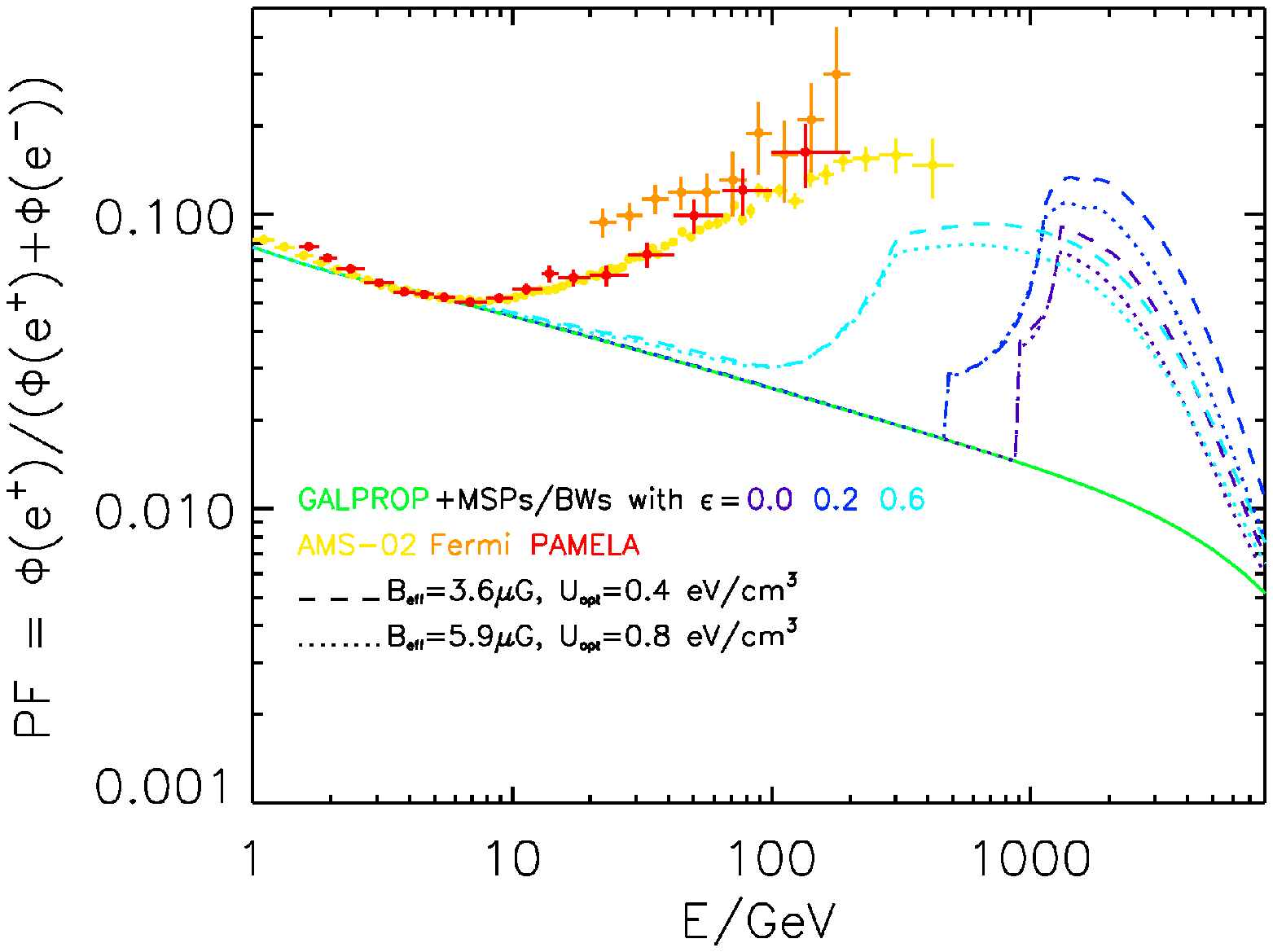}
\caption{Same as Figure~\ref{fig:PF1}, but for $\kappa_0 = 0.1$~kpc$^2$\,Myr$^{-1}$ and $\eta_{\rm p,max} = 0.3.$}\label{fig:PF2}
\end{figure}

\begin{figure}
\epsscale{1.0}
\plotone{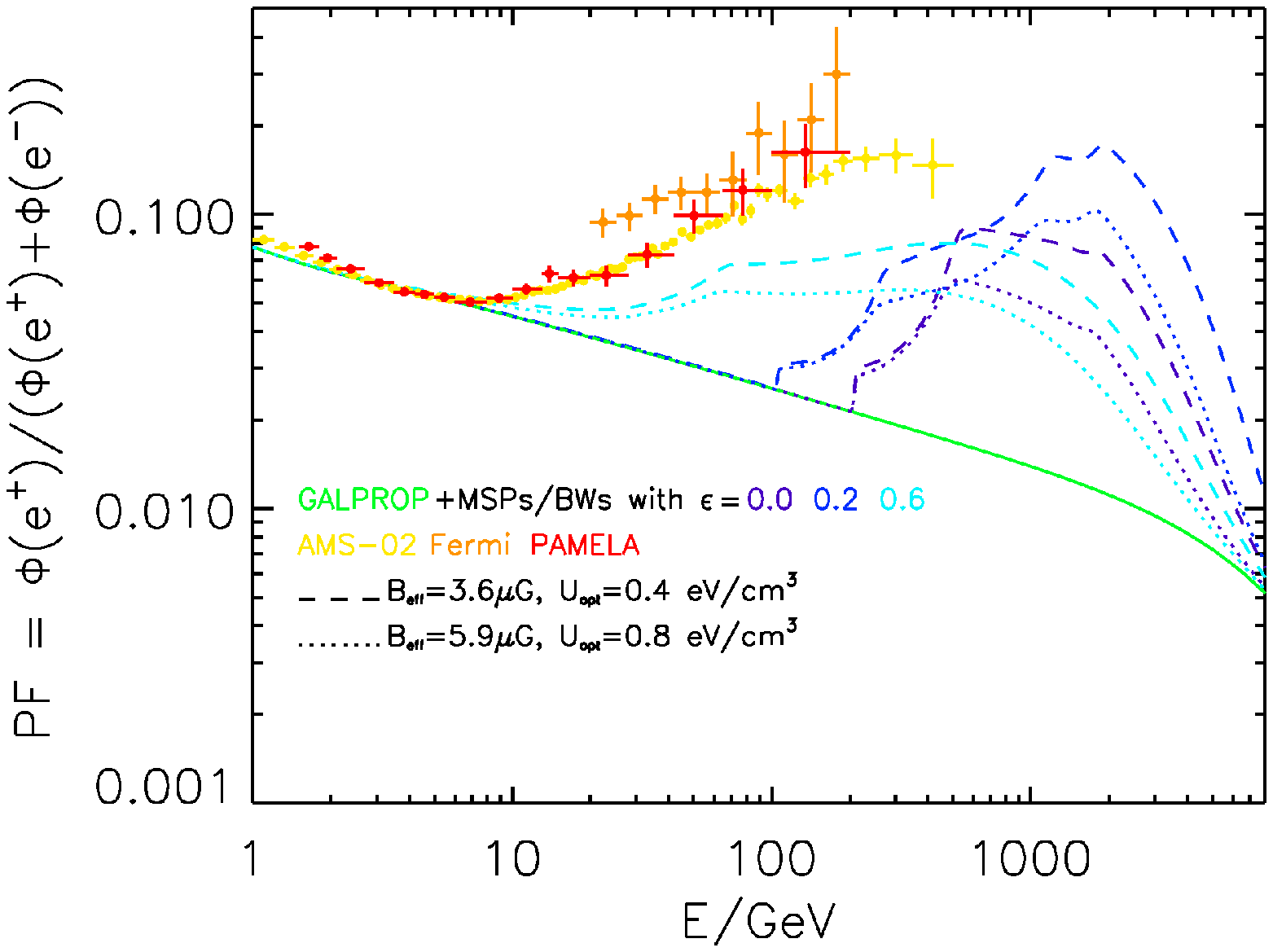}
\caption{Same as Figure~\ref{fig:PF1}, but for $\kappa_0 = 0.01$~kpc$^2$\,Myr$^{-1}$ and $\eta_{\rm p,max} = 0.1.$}\label{fig:PF3}
\end{figure}

\begin{figure}
\epsscale{1.0}
\plotone{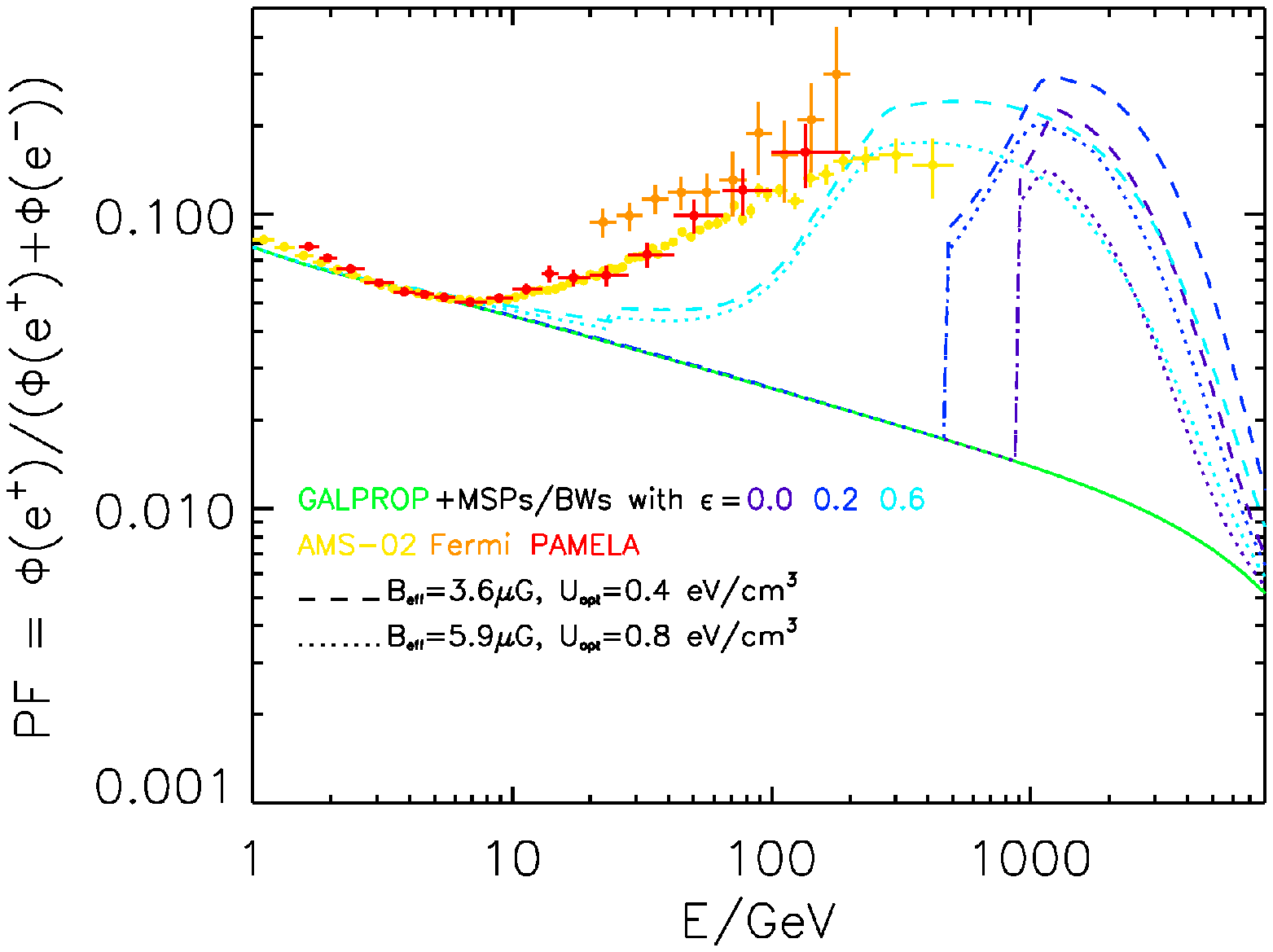}
\caption{Same as Figure~\ref{fig:PF1}, but for $\kappa_0 = 0.01$~kpc$^2$\,Myr$^{-1}$ and $\eta_{\rm p,max} = 0.3.$}\label{fig:PF4}
\end{figure}

\begin{figure}
\epsscale{1.0}
\plotone{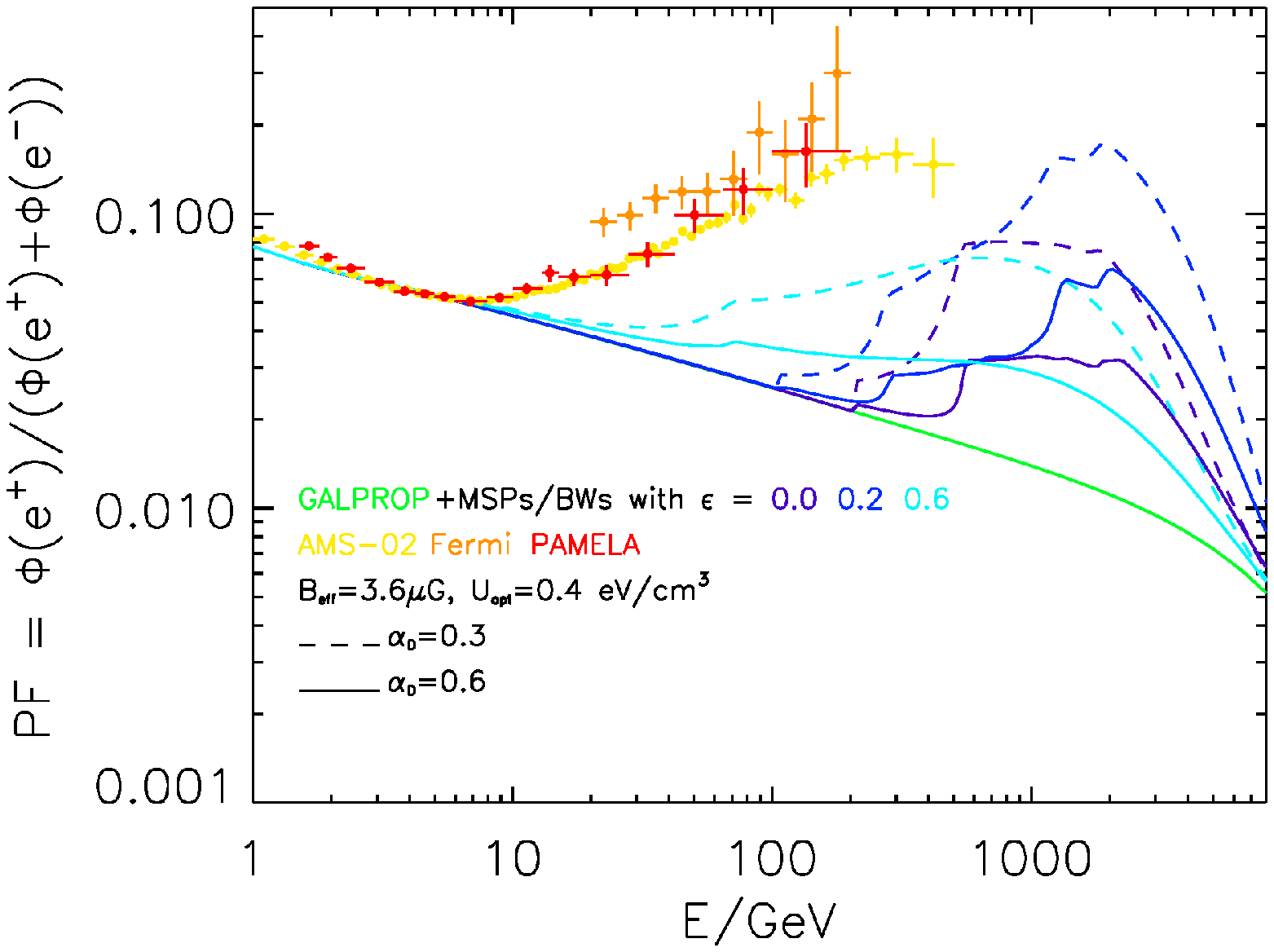}
\caption{Same as Figure~\ref{fig:PF1}, but for $B_{\rm eff} = 3.6\,\mu$G, $U_{\rm opt}$ = 0.4 eV\,cm$^{-3}$, $\eta_{\rm p,max} = 0.1$,  $\kappa_0 = 0.1$~kpc$^2$\,Myr$^{-1}$, and $\alpha_{\rm D} = 0.3$ and $0.6$.}\label{fig:PF4a}
\end{figure}

\begin{figure}
\epsscale{1.0}
\plotone{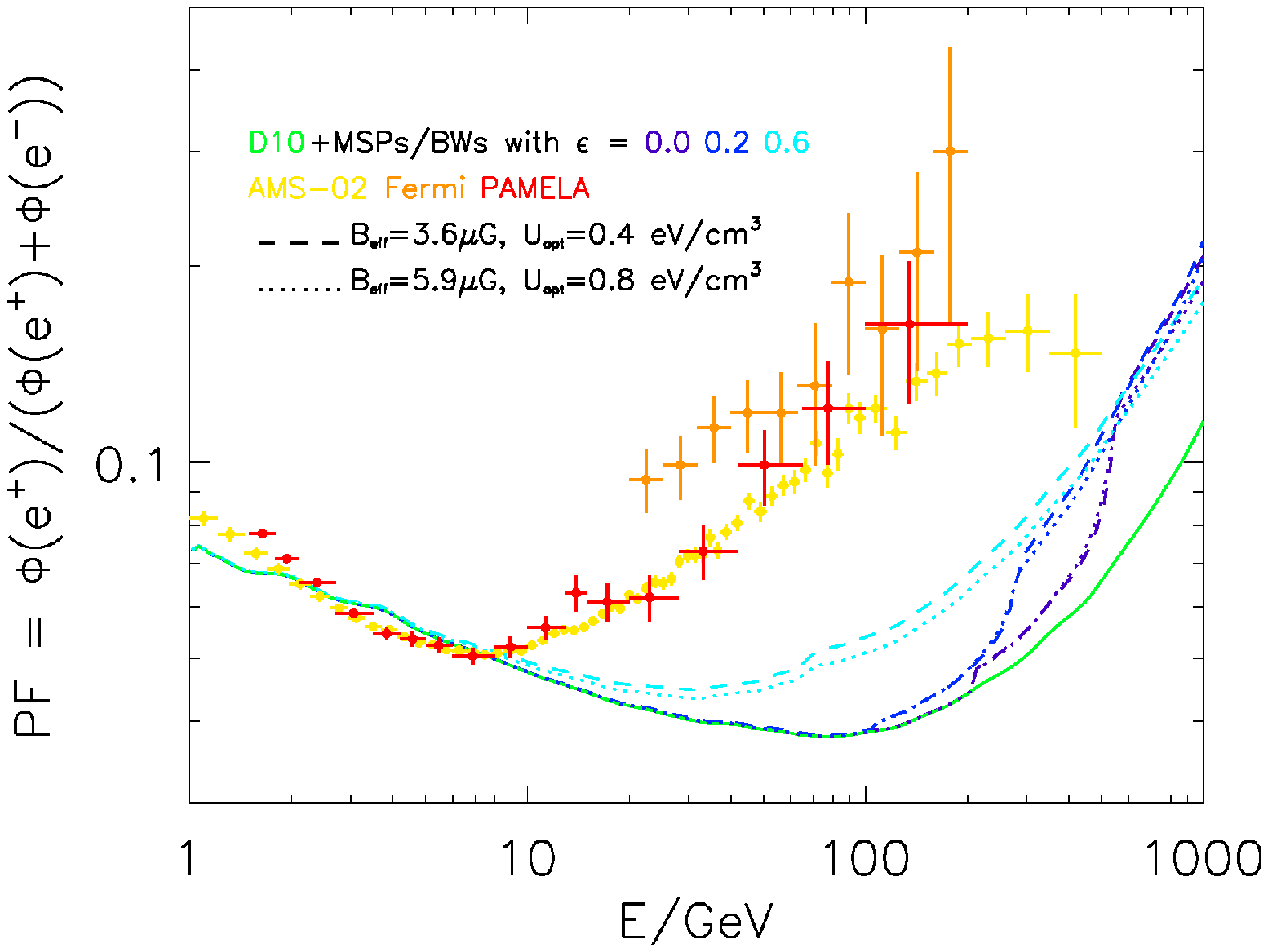}
\caption{Same as Figure~\ref{fig:PF1}, but using the background model of D10, for $\kappa_0 = 0.1$~kpc$^2$\,Myr$^{-1}$ and $\eta_{\rm p,max} = 0.1$.}\label{fig:PF5}
\end{figure}

\begin{figure}
\epsscale{1.0}
\plotone{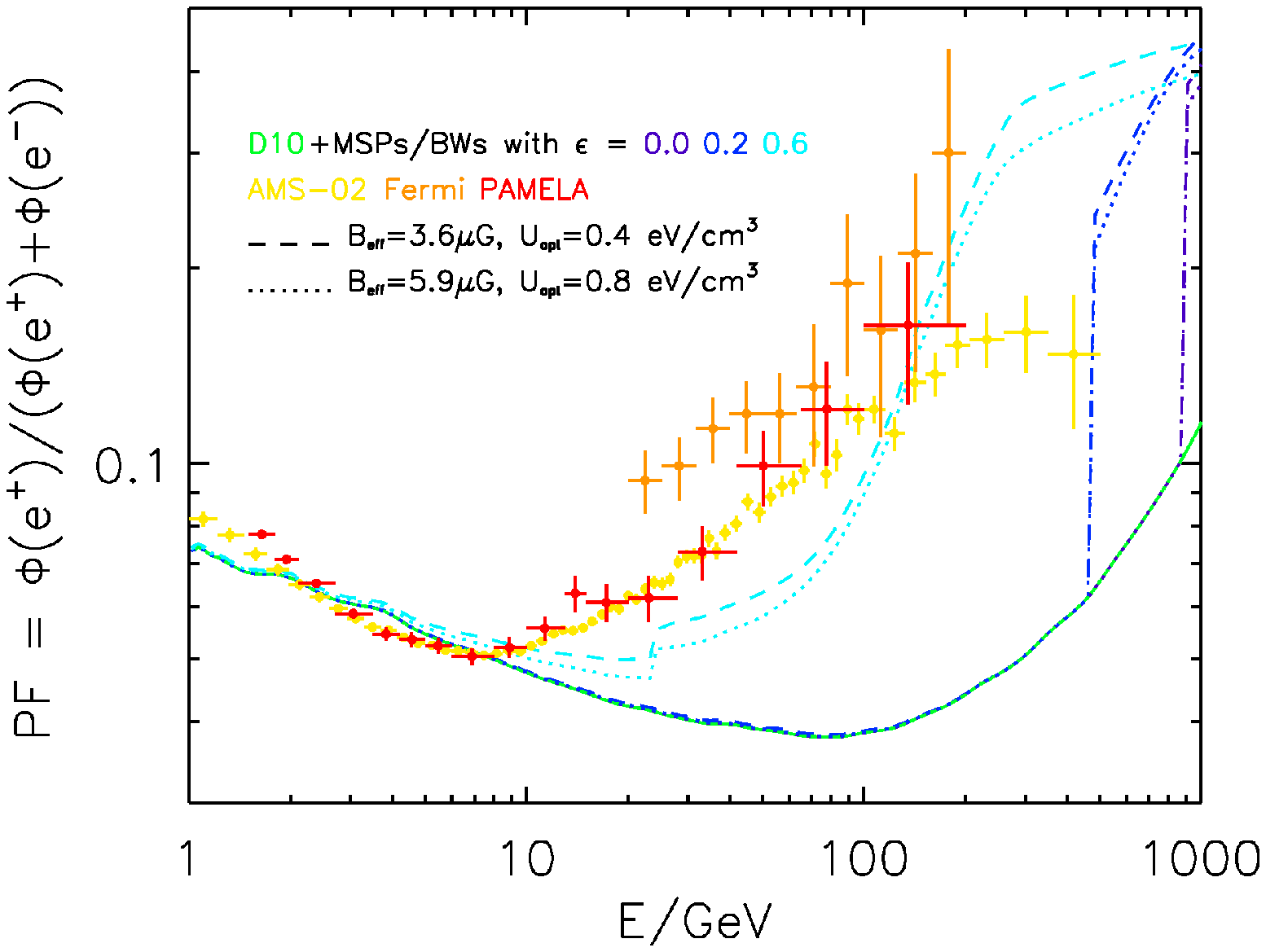}
\caption{Same as Figure~\ref{fig:PF1}, but using the background model of D10, for $\kappa_0 = 0.01$~kpc$^2$\,Myr$^{-1}$ and $\eta_{\rm p,max} = 0.3$.}\label{fig:PF6}
\end{figure}

Figure~\ref{fig:PF1} shows the measured PF \citep[e.g.,][]{Accardo14} as well as the GALPROP and synthesis plus BW / RB contributions, for $\kappa_0 = 0.1$~kpc$^2$\,Myr$^{-1}$ and $\eta_{\rm p,max} = 0.1$. The largest contribution is found $\sim100$~GeV in the case of $\varepsilon=0.6$ and $B=3.6\,\mu$G. Figure~\ref{fig:PF2} is the same, but for $\kappa_0 = 0.1$~kpc$^2$\,Myr$^{-1}$, $\eta_{\rm p,max}=0.3$, while in Figure~\ref{fig:PF3} we use $\kappa_0 = 0.01$~kpc$^2$\,Myr$^{-1}$, $\eta_{\rm p,max}=0.1$, and in Figure~\ref{fig:PF4}, $\kappa_0 = 0.01$~kpc$^2$\,Myr$^{-1}$, $\eta_{\rm p,max}=0.3$. In Figure~\ref{fig:PF4a}, we show results for different choices of $\alpha_{\rm D}$, the highest ratio (above 1~TeV) occurring for the lowest value of $\alpha_{\rm D}$. Figure~\ref{fig:PF5} is for the background model of D10, for $\kappa_0 = 0.1$~kpc$^2$\,Myr$^{-1}$, $\eta_{\rm p,max}=0.1$, while Figure~\ref{fig:PF6} is for the background model of D10, for $\kappa_0 = 0.01$~kpc$^2$\,Myr$^{-1}$, $\eta_{\rm p,max}=0.3$. We note that the BW / RB component makes a significant contribution at a few hundred GeV, increasing with $\eta_{\rm p,max}$ and decreasing with $\kappa_0$, while the result is very sensitive to the choice of background model. Some parameter combinations are excluded by the data, e.g., $\kappa_0 = 0.01$~kpc$^2$\,Myr$^{-1}$, $\eta_{\rm p,max}=0.3$, and $\varepsilon = 0.6$, depending on the choice of background model.


\section{Discussion}
\label{sec:discuss}
Our results have shown that for certain ranges of parameters, MSPs could make a significant contribution to the local cosmic-ray lepton spectrum.  On the other hand, our calculations show that some parameter combinations can also be ruled out.  Shock-accelerated positrons and electrons from BWs and RBs make a much larger contribution than the rest of the MSP population (above $\sim100$~GeV), since a much higher fraction of the pulsars' spin-down power goes into particle power in these sources (compared to the pair cascades). The BW and RB contribution nearly reaches the observed positron fraction for $\kappa_0 = 0.01$~kpc$^2$\,Myr$^{-1}$, $\eta_{\rm p,max}=0.1$, $\varepsilon=0.6$ and $B_{\rm eff}=3.6\,\mu$G. However, the extreme parameter combination $\kappa_0 = 0.01$~kpc$^2$\,Myr$^{-1}$, $\eta_{\rm p,max}=0.3$ and  $\varepsilon=0.6$ can be ruled out, while $\kappa_0 = 0.01$~kpc$^2$\,Myr$^{-1}$, $\eta_{\rm p,max}=0.1$ and $\varepsilon=0.2$ would predict a rise in both the electron and positron spectra and the positron fraction above 500~GeV. We also note that the uncertainty in the energy dependence of the spatial diffusion coefficient (and background model) may lead to large uncertainties in the flux predictions, especially at very high energies. Future measurements by \textit{AMS-02} extending the spectra to larger energies may help constrain even more of parameter space. We also note that the level of the pair spectra starts to saturate for larger $\varepsilon$. This is because more photons convert to pairs at lower energies in this case (mostly due to the lower curvature radius of the distorted field lines). One therefore cannot increase the source flux without bounds by increasing $\varepsilon$, since there is a limit to the maximum contribution one would obtain as a function of $\varepsilon$. We have chosen values for $\varepsilon$ that are reasonable, simulating the range of offsets obtained in newer solutions to realistic magnetospheres \citep{HM11b}. 

We found that the PF could be nicely reproduced if the population synthesis component were shifted to an energy higher by a factor of $\sim4$ and at a level higher by a factor of $\sim10$. Uncertainties in the transport and source properties may account for some of this shortfall, so that these numbers are not too large. However, we rather interpret this as pointing to the fact that the synthesis component has the correct spectral shape to explain the data (i.e., hard enough spectral index; see, e.g., \citealt{Gaggero14}), given the assumed background model.

We have taken the approach of computing the positron contribution from MSPs using reasonable parameter values rather than tuning the parameters just to fit the PF. This is because we do not expect the MSPs to explain all of the data, since there are many good arguments why young pulsars in PWNe, as well as SNRs, may make even larger contributions. The rise in PF may plausibly be the result of contributions by many sources \citep[e.g.,][]{DiMauro14}. One should therefore be careful not to overproduce the data by not considering the cumulative contribution from all viable sources. The argument may actually be reversed: if one has access to solid predictions for the contribution of several sources, one could in principle constrain parameters such as the shock acceleration efficiency so as not to overproduce the observed flux. In practice, this may be difficult, though, given the large number of free parameters and model uncertainties. 

It is clear that the predicted spectrum of secondary electrons and positrons from cosmic ray interactions in the ISM has a significant influence on the rise and shape of the PF, and therefore the need for a primary positron contribution. \citet{Blum13} argue that  the PF fraction rise can even be explained exclusively by secondary positrons.  Such a hypothesis would imply strong constraints on the properties of the primary positron sources, including the roughly 50,000 Galactic MSPs, shock acceleration in BW and RB binaries, and young PWNe. However, the calculations of \citet{Blum13} have a factor $\sim2$ uncertainty, implying that there may be a primary contribution of equal strength around several hundred GeV, and even at a higher relative level at lower energies. Furthermore, a source that cuts off at lower energies will not violate this upper limit. We have also noted in Section~\ref{sec:intro} the conclusion by \citet{Dado15} that a primary source of positrons is strongly required to explain the cosmic-ray data. We therefore envision room for both a primary and secondary contribution to the observed excess.

It became clear during our study that the results are very dependent on the properties of the closest few sources, since they dominate any contribution from the larger, more distant, population. A change in the properties of these nearby sources for a new realization of the synthesis population may therefore impact our predictions. To investigate this matter, we split the population into two parts, and compared the cumulative electron and positron spectra from these two subpopulations. While there were some differences, the effect was fortunately rather minor, and we can therefore have some confidence in our predictions.

Our population synthesis uses radio survey sensitivity and \textit{Fermi} three-year point source sensitivity maps, normalizing to the number of detected radio MSPs from those surveys only, and to the detected $\gamma$-ray MSPs in the Second Pulsar Catalog \citep{Abdo13}, all of which are radio-loud. The MSPs discovered in radio followup observations of unidentified \textit{Fermi} sources are included in our simulated population of MSPs not detected by radio surveys but detected by \textit{Fermi} as point sources. However, it is possible that there is a contribution to the cosmic-ray flux from BWs and RBs very close to Earth that, due to difficulty of radio detection in these eclipsing and/or  obscured systems, have not yet been identified. Future detections will impact the normalization of the output from the synthesis component, and may also enhance the flux prediction from the BW / RB component. Our predicted contribution from the latter should therefore be seen as a lower limit, since we have only considered detected BWs and RBs in our calculation.

We note that shock acceleration in binary systems should lead to non-thermal emission that is modulated at the orbital period \citep{Arons93,Bogdanov11}. 
None has been seen in the $\gamma$-ray band so far, although there have been some detections of X-ray emission modulated at the orbital periods in the BW systems B1957+20 \citep{HB12}, J2215+5135, and J2256-1024 \citep{Gentile14}. Future detections of such high-energy modulated signals will provide further confirmation and constraints on the shock acceleration scenario.

\citet{Kisaka12} suggested that pair-starved MSPs may be responsible for a large peak in the total electron spectrum at $10-100$~TeV, and that non-pair-starved MSPs with multiplicities of $\sim2~000$ may contribute significantly (near 100\%) to the PF above 10~GeV. There are, however, a number of differences in our respective approaches. \citet{Kisaka12} used fixed values for $P$ and $B_{\rm s}$ for all members in their population. They furthermore assumed energy equipartition between the particles and the magnetic field, which seems to imply a conversion efficiency (from spin-down luminosity to particle power) of $\eta\sim 50\%$, while we find $\eta\sim1\%$ from our pair cascade modeling. They also assume a lower average Galactic magnetic field ($B = 1\,\mu$G). Finally, they integrate the injected spectra over the age of the MSPs while we follow a steady-state approach. Most if not all of these differences should lead to an enhanced particle flux in their case.

It has been suggested that isotropy may be a discriminator between an astrophysical and dark matter origin of the rise in PF \citep[e.g.,][]{Buesching08,Aguilar14}. However, if several nearby sources contribute, any potential anisotropy may be washed out \citep[see also][]{Feng15}. \citet{Linden13} also raised some issues, e.g., inhomogeneous magnetic fields, diffusion properties beyond the standard assumptions, or pulsar proper motion. Indeed, \citet{Kistler12} noted that turbulence in the local magnetic field may lead to filamentary structures or ``streams'' of cosmic rays. These streams may redirect or wash out signatures of local sources  that may otherwise have contributed under the assumption of isotropy (i.e., making the terrestrial spectra nearly featureless), or conversely lead to an enhancement in the contribution by otherwise negligible sources by concentrating their fluxes. One should lastly consider the ``coherence length'' or mean free path beyond which any anisotropy would disappear. Given these uncertainties, we would argue that anisotropy measurements may not be such a clear discriminator after all.


\section{Conclusion}
\label{sec:concl}
In this paper, we carefully assessed the contribution of MSPs to the cosmic-ray lepton spectra at Earth using a population synthesis code and a pair cascade code to calculate realistic source spectra. We also considered the contribution of binary BW / RB systems, which may further accelerate pairs escaping from the MSP magnetospheres in intrabinary shocks. 

We find that the predicted MSP particle flux increases for non-zero magnetic field offset parameters $\varepsilon$. This is expected, since a larger value for the offset of the surface magnetic field with respect to the non-perturbed magnetic axis leads to an increase in the acceleration potential for some regions in azimuthal phase (and a decrease in others). This in turn results in an enhancement in both the number of particles (since the multiplicity will be higher) as well as the maximum particle source energy (given a larger local electric field in some regions). We find that the MSPs from the synthesis model make only a modest contribution to the terrestrial cosmic-ray flux at a few tens of GeV, after which this spectral component cuts off. This is because the maximal injected particle energy is limited by a maximal electric field, which depends on the MSP source properties such as $P$, $B_{\rm s},$ and $\varepsilon$. The effect of different Galactic magnetic fields and soft-photon energy densities is also shown: an increased field and densities lead to increased energy losses, and vice versa. We have bracketed these losses, and note that the flux uncertainty is not too large, given these uncertainties in magnetic field and energy densities.  Although the PF is somewhat enhanced above $\sim10$~GeV, our added MSP synthesis component fails to reproduce the high-energy rise for the parameters considered. 

The BW / RB component contributes more substantially above several hundred GeV, given the fact that they further accelerate the electron-positron pairs in their strong intrabinary shocks. For some parameter combinations, this component may even exceed the measured positron spectrum, and may violate the PF at high energies, depending on the background model. 

Alternative sources of primary positrons such as young, nearby pulsars or SNRs should also contribute to the cosmic-ray electron and positron flux. Future observations and modeling should continue to constrain the properties of these source classes, as well as improve our understanding of Galactic structure and particles within our Galaxy.

\acknowledgments
CV is supported by the South African National Research Foundation. AKH and PLG acknowledge support from the NASA Astrophysics Theory Program. We thank Julie McEnery, Chuck Dermer, Kent Wood, Marius Potgieter, Driaan Bisschoff (and also acknowledge the late Okkie de Jager) for stimulating discussions.

\clearpage

\clearpage

\clearpage

\clearpage

\clearpage

\clearpage

\clearpage

\clearpage

\clearpage


\clearpage

\clearpage

\clearpage

\clearpage

\clearpage

\clearpage

\clearpage

\clearpage

\clearpage

\clearpage

\clearpage

\clearpage

\clearpage

\clearpage
\begin{deluxetable}{lcccccccccc}
\tabletypesize{\footnotesize}
\tablecaption{Measured and derived parameters of BW pulsars.\label{tab1}}
\tablenum{1}
\tablehead{\colhead{Name} & \colhead{$P_{\rm ms}$} & \colhead{$\dot{P}_i$} & \colhead{$L_{\rm sd}$\tablenotemark{a}} & \colhead{$B_8$\tablenotemark{b}} & \colhead{$d$} & \colhead{$P_{\rm b}$} & \colhead{$M_{\rm comp}$} & \colhead{$a_{11}$} & \colhead{$E_{\rm cut}$} & \colhead{Ref.} \\ 
\colhead{} & \colhead{} & \colhead{$\left(10^{-20}\right)$} & \colhead{$\left(10^{34}~{\rm  erg\,s}^{-1}\right)$} & \colhead{} & \colhead{(kpc)} & \colhead{(h)} & \colhead{($M_\odot$)} & \colhead{} & \colhead{(TeV)} & \colhead{}} 

\startdata
J0023+0923\tablenotemark{c} & 3.05 & 1.15 & 2.50 & 4.88 & 0.7 & 3.3 & 0.016 & 1.01 & 2.40 & 1 \\
J0610$-$2100\tablenotemark{c} & 3.86 & 0.34 & 0.36 & 2.96 & 3.5 & 6.9 & 0.025 & 1.65 & 3.04 & 2 \\
J1124$-$3653\tablenotemark{c} & 2.41 & 0.57 & 2.50 & 3.05 & 1.7 & 5.4 & 0.027 & 1.40 & 2.03 & 1 \\
J1301+0833\tablenotemark{c} & 1.84 & 0.95 & 9.36 & 3.44 & 0.7 & 6.5 & 0.024 & 1.59 & 1.37 & 3 \\
J1311$-$3430\tablenotemark{c} & 2.56 & 2.08 & 7.64 & 6.01 & 1.4 & 1.56 & 0.008 & 0.61 & 2.33 & 4 \\
J1446$-$4701\tablenotemark{c} & 2.19 & 1.01 & 5.93 & 3.88 & 1.5 & 6.7 & 0.019 & 1.62 & 1.52 & 5 \\
J1544+4937\tablenotemark{c} & 2.16 & 0.31 & 1.87 & 2.12 & 1.2 & 2.8 & 0.018 & 0.91 & 2.72 & 6 \\
J1731$-$1847                & 2.34 & 2.47 & 11.9 & 6.26 & 2.5 & 7.5 & 0.04 & 1.75 & 1.23 & 7 \\
J1745+1017\tablenotemark{c} & 2.65 & 0.23 & 0.75 & 2.02 & 1.36 & 17.5 & 0.016 & 3.07 & 1.86 & 8 \\
J1810+1744\tablenotemark{c} & 1.66 & 0.45 & 6.08 & 2.26 & 2 & 3.6 & 0.044 & 1.07 & 1.86 & 1 \\
J1959+2048\tablenotemark{c} & 1.61 & 0.72 & 10.6 & 2.80 & 1.53 & 9.2 & 0.021 & 2.00 & 1.19 & 9 \\
J2047+1053\tablenotemark{c} & 4.29 & 2.00 & 1.56 & 7.63 & 2 & 3 & 0.035 & 0.95 & 2.78 & 3 \\
J2051$-$0827\tablenotemark{c} & 4.51 & 1.23 & 0.83 & 6.14 & 1 & 2.4 & 0.027 & 0.82 & 3.51 & 2 \\
J2214+3000\tablenotemark{c} & 3.12 & 1.46 & 2.96 & 5.57 & 1.32 & 10 & 0.014 & 2.11 & 1.59 & 10, 11 \\
J2234+0944\tablenotemark{c} & 3.63 & 1.94 & 2.50 & 6.91 & 1 & 10 & 0.015 & 2.11 & 1.66 & 3, 5 \\
J2241$-$5236\tablenotemark{c} & 2.19 & 0.67 & 3.90 & 3.15 & 0.5 & 3.4 & 0.012 & 1.03 & 2.12 & 12 \\
J2256$-$1024\tablenotemark{c} & 2.29 & 1.58 & 8.11 & 4.96 & 0.6 & 5.1 & 0.034 & 1.35 & 1.54 & 1 \\
\enddata

\tablenotetext{a}{We have used the equation of state (EOS) described in Section~\ref{sec:BW}. For canonical values, divide by a factor 1.56.}
\tablenotetext{b}{We have used the EOS described in Section~\ref{sec:BW}. For canonical values, multiply by a factor 0.78.}
\tablenotetext{c}{\textit{Fermi} LAT pulsations have been seen from this pulsar.}

\tablecomments{The columns are as follows: Pulsar name; pulsar period in milliseconds; intrinsic (Shklovskii-corrected) period derivative, as calculated from the announced spin-down luminosities; spin-down luminosity; surface magnetic field in units of $10^8$~G; distance; binary period; companion mass; binary separation in units of $10^{11}$~cm; spectral cutoff energy. We assume a pulsar radius of $R=9.9\times10^5$~cm and moment of inertia of $I=1.56\times10^{45}$~g\,cm$^2$.}
\tablerefs{(1)~\citet{Hessels11}; (2)~\citet{Espinoza13}; (3)~\citet{Ray12}; (4)~\citet{Pletsch12}; (5)~\citet{Keith12}; (6)~\citet{Bhattacharyya13}; (7)~\citet{Keith10}; (8)~\citet{Barr13}; (9)~\citet{Fruchter90}; (10)~\citet{Roberts11}; (11)~\citet{Ransom10}.}
\end{deluxetable}

\clearpage
\begin{deluxetable}{lcccccccccc}
\tabletypesize{\footnotesize}
\tablecaption{Measured and derived parameters of RB pulsars.\label{tab2}}
\tablenum{2}
\tablehead{\colhead{Name} & \colhead{$P_{\rm ms}$} & \colhead{$\dot{P}_i$} & \colhead{$L_{\rm sd}$\tablenotemark{a}} & \colhead{$B_8$\tablenotemark{b}} & \colhead{$d$} & \colhead{$P_{\rm b}$} & \colhead{$M_{\rm comp}$} & \colhead{$a_{11}$} & \colhead{$E_{\rm cut}$} & \colhead{Ref.} \\ 
\colhead{} & \colhead{} & \colhead{$\left(10^{-20}\right)$} & \colhead{$\left(10^{34}~{\rm  erg\,s}^{-1}\right)$} & \colhead{} & \colhead{(kpc)} & \colhead{(h)} & \colhead{($M_\odot$)} & \colhead{} & \colhead{(TeV)} & \colhead{}} 

\startdata
J1023+0038                    & 1.69 & 1.20 & 15.4 & 3.72 & 0.6 & 4.8 & 0.2 & 1.33 & 1.33 & 1 \\
J1628$-$3205                  & 3.21 & 1.13 & 2.11 & 4.96 & 1.2 & 5 & 0.16 & 1.36 & 2.15 & 2 \\
J1723$-$2837                  & 1.86 & 0.75 & 7.18 & 3.08 & 0.75 & 14.8 & 0.4 & 2.90 & 1.09 & 3, 4 \\
J1816+4510\tablenotemark{c}   & 3.19 & 4.03 & 7.64 & 9.34 & 2.4 & 8.7 & 0.16 & 1.97 & 1.30 & 5 \\
J2129$-$0429                  & 7.61 & 43.54 & 6.08 & 47.4 & 0.9 & 15.2 & 0.37 & 2.94 & 1.12 & 6 \\
J2215+5135\tablenotemark{c}   & 2.61 & 2.79 & 9.67 & 7.03 & 3 & 4.2 & 0.22 & 1.22 & 1.55 & 6 \\
J2339$-$0533\tablenotemark{c} & 2.88 & 1.39 & 3.59 & 5.21 & 0.4 & 4.6 & 0.26 & 1.30 & 1.93 & 7, 8 \\
\enddata

\tablenotetext{a}{For canonical values, divide by a factor 1.56.}
\tablenotetext{b}{For canonical values, multiply by a factor 0.78.}
\tablenotetext{c}{\textit{Fermi} LAT pulsations have been seen from this pulsar.}
\tablecomments{The columns are the same as for Table~\ref{tab1}.}

\tablerefs{(1)~\citet{Archibald09}; (2)~\citet{Ray12}; (3)~\citet{Roberts11}; (4)~\citet{Crawford13}; (5)~\citet{Kaplan12}; (6)~\citet{Hessels11}; (7)~\citet{Kong12}; (8)~\citet{Ray14}.}
\end{deluxetable}


\begin{thebibliography}{}
\bibitem[Abdo et al.(2013)]{Abdo13} Abdo, A.~A. et al. 2013, \apjs, 208, 17
\bibitem[Accardo et al.(2014)]{Accardo14} Accardo, L. et al. 2014, \prl, 113, 121101
\bibitem[Ackermann et al.(2012)]{Ackermann12} Ackermann, M. et al. 2012, \prl, 108, 011103
\bibitem[Adriani et al.(2009)]{Adriani09} Adriani, O. et al. 2009, \nat, 458, 607
\bibitem[Adriani et al.(2013)]{Adriani13} Adriani, O. et al. 2013, \prl, 111, 081102
\bibitem[Aguilar et al.(2013)]{Aguilar13} Aguilar, M. et al. 2013, \prl, 110, 141102
\bibitem[Aguilar et al.(2014)]{Aguilar14} Aguilar, M. et al. 2014, \prl, 113, 121102
\bibitem[Aharonian et al.(1995)]{Aharonian95} Aharonian, F.~A., Atoyan, A.~M., \& Voelk, H.~J. 1995, \aap, 294, L41
\bibitem[Alpar et al.(1982)]{Alpar82} Alpar, M.~A., Cheng, A.~F., Ruderman, M.~A., \& Shaham, J. 1982, \nat, 300, 728
\bibitem[Amano \& Kirk(2013)]{AK13} Amano, T. \& Kirk, J.~G. 2013, ApJ, 770, 18
\bibitem[Archibald et al.(2009)]{Archibald09} Archibald, A.~M. et al. 2009, Science, 324, 1411
\bibitem[Arons \& Scharlemann(1979)]{AS79} Arons, J. \& Scharlemann, E.~T. 1979, \apj, 231, 854
\bibitem[Arons(1981)]{Arons81} Arons, J. 1981, in Origin of Cosmic Rays, IAU Symposium, ed.\ G. Setti, G. Spada, \& A.~W.\ Wolfendale, 94, 175
\bibitem[Arons(1996)]{Arons96} Arons, J. 1996, \aaps, 120, C49
\bibitem[Arons \& Tavani(1993)]{Arons93} Arons, J., \& Tavani, M. 1993, \apj, 403, 249
\bibitem[Barr et al.(2013)]{Barr13} Barr, E.~D. et al. 2013, \mnras, 429, 1633
\bibitem[Beck(2009)]{Beck09} Beck, R. 2009, Astrophys.\ Space Sci.\ Trans., 5, 43
\bibitem[Bhattacharyya et al.(2013)]{Bhattacharyya13} Bhattacharyya, B. et al. 2013, \apjl, 773, L12
\bibitem[Blasi(2009)]{Blasi09} Blasi, P. 2009, \prl, 103, 051104
\bibitem[Blasi \& Amato(2011)]{Blasi11} Blasi, P., \& Amato, E. 2011, in High-Energy Emission from Pulsars and their Systems, ed. D.~F. Torres, \& N. Rea, 624
\bibitem[Blies \& Schlickeiser(2012)]{Blies12} Blies, P., \& Schlickeiser, R. 2012, \apj, 751, 71
\bibitem[Blum et al.(2013)]{Blum13} Blum, K., Katz, B., \& Waxman, E. 2013, \prl, 111, 211101
\bibitem[Blumenthal \& Gould(1970)]{Blumenthal70} Blumenthal, G.~R., \& Gould, R.J. 1970, Rev.\ Mod.\ Phys., 42, 237
\bibitem[Bogdanov et al.(2007)]{Bogdanov07} Bogdanov, S., Rybicki, G. B. \& Grindlay, J. E. 2007, \apj, 670, 668
\bibitem[Bogdanov \& Grindlay(2009)]{BogdanovGrindlay09} Bogdanov, S. \& Grindlay, J. E. 2009, \apj, 703, 1557
\bibitem[Bogdanov et al.(2011)]{Bogdanov11} Bogdanov, S. et al. 2011, \apj, 742, 97
\bibitem[Bogdanov(2013)]{Bogdanov13} Bogdanov, S. 2013, \apj, 762, 96
\bibitem[Boulares(1989)]{Boulares89} Boulares, A., 1989, \apj, 342, 807
\bibitem[Bucciantini et al.(2011)]{Bucc11} Bucciantini, N., Arons, J \& Amato, E. 2011, \mnras, 410, 381
\bibitem[Burlaga \& Ness(2014)]{Burlaga14} Burlaga, L.~F. \& Ness, N.~F. 2014, \apj, 784, 146
\bibitem[B\"usching et al.(2008a)]{Buesching08} B{\"u}sching, I., Venter, C., \& de Jager, O.~C. 2008a, Adv.\ Space Res., 42, 497
\bibitem[B\"usching et al.(2008b)]{Buesching08b} B{\"u}sching, I., de Jager, O.~C., Potgieter, M.~S., \& Venter, C. 2008b, \apj, 678, L39
\bibitem[Chi et al.(1996)]{Chi96} Chi, X., Cheng, K.~S., Young, E.~C.~M. 1996, \apjl, 459, L83
\bibitem[Contopoulos et al.(2014)]{Contopoulos14} Contopoulos, I., Kalapotharakos, C., \& Kazanas, D. 2014, \apj, 781, 46
\bibitem[Cowsik \& Burch(2010)]{Cowsik10} Cowsik, R., \& Burch, B. 2010, \prd, 82, 023009
\bibitem[Crawford et al.(2013)]{Crawford13} Crawford, F. et al. 2013, \apj, 776, 20
\bibitem[Dado \& Dar(2015)]{Dado15} Dado, S., \& Dar, A. 2015, arXiv:1504.03261
\bibitem[Daugherty \& Harding(1982)]{DH82} Daugherty, J.~K. \& Harding, A.~K. 1982, \apj, 252, 337
\bibitem[Daugherty \& Harding(1983)]{DH83} Daugherty, J.~K., \& Harding, A.~K. 1983, \apj, 273, 761
\bibitem[de Jager et al.(1996)]{DeJager96} de Jager, O. C., {Harding}, A. K., {Michelson}, P.~F., {Nolan}, P. L., {Sreekumar}, P. \& {Thompson}, D.~J. 1996, \apj, 457, 253
\bibitem[Delahaye et al.(2010)]{D10} Delahaye, T., Lavalle, J., Lineros, R., Donato, F., \& Fornengo, N. 2010, \aap, 524, A51 (D10)
\bibitem[Demorest et al.(2010)]{Demorest10} Demorest, P.~B., Pennucci, T., Ransom, S.~M., Roberts, M.~S.~E., \& Hessels, J.~W.~T. 2010, \nat, 467, 1081
\bibitem[Deutsch(1955)]{Deutsch55} Deutsch, A.~J. 1955, Annales d'Astrophysique, 18, 1
\bibitem[Di Mauro et al.(2014)]{DiMauro14} Di Mauro, M., Donato, F., Fornengo, N., Lineros, R.,  \& Vittino, A. 2014, \jcap, 4, 6
\bibitem[Dogiel \& Sharov(1990)]{Dogiel90} Dogiel, A.~V., \& Sharov, S.~G. 1990, Proc.\ ICRC, 4, 109
\bibitem[Dyks \& Harding(2004)]{Dyks04} Dyks, J. \& Harding, A.~K. 2004, \apj, 614, 869
\bibitem[Erber(1966)]{Erber66} Erber, T. 1966, Rev.\ Mod.\ Phys., 38, 626
\bibitem[Espinoza et al.(2013)]{Espinoza13} Espinoza, C.~M. et al. 2013, \mnras, 430, 571
\bibitem[Fan et al.(2010)]{Fan10} Fan, Y.-Z., Zhang, B., \& Chang, J. 2010, Int.\ J.\ Mod.\ Phys.\ D, 19, 2011
\bibitem[Feng \& Zhang(2015)]{Feng15} Feng, J., \& Zhang, H.-H. 2015, arXiv:1504.03312
\bibitem[Fruchter et al.(1990)]{Fruchter90} Fruchter, A.~S. et al. 1990, \apj, 351, 642
\bibitem[Gaggero et al.(2014)]{Gaggero14}  Gaggero, D., Maccione, L., Grasso, D., Di Bernardo, G., \& Evoli, C. 2014, \prd, 89, 083007
\bibitem[Gendelev et al.(2010)]{Gendelev10} Gendelev, L., Profumo, S., \& Dormody, M. 2010, \jcap, 2, 16
\bibitem[Genolini et al.(2015)]{Genolini15} Genolini, Y., Putze, A., Salati, P., Serpico, P.~D. 2015, submitted to \aap (arXiv:1504.03134)
\bibitem[Gentile et al.(2014)]{Gentile14} Gentile, P. et al. 2014, \apj, 783, 69
\bibitem[Goldreich \& Julian(1969)]{Goldreich69} Goldreich, P., \& Julian, W.~H. 1969, \apj, 157, 869
\bibitem[Gonthier et al.(2015)]{Gonthier15} Gonthier, P.~L. et al., in prep.
\bibitem[Grasso et al.(2009)]{Grasso09} Grasso, D. et al. 2009, Astropart.\ Phys., 32, 140
\bibitem[Grimani(2007)]{Grimani07} Grimani, C. 2007, \aap, 474, 339
\bibitem[Gupta \& Torres(2014)]{Gupta14} Gupta, N. \& Torres, D.~F. 2014, \mnras, 441, 3122
\bibitem[Han \& Qiao(1994)]{HQ94} Han, J.~L., \& Qiao, G.~J. 1994, \aap, 288, 759
\bibitem[Han et al.(2002)]{Han02} Han, J.~L., Manchester, R.~N., Lyne, A.~G., \& Qiao, G.~J. 2002, \apj, 570, L17
\bibitem[Han et al.(2006)]{Han06} Han, J.~L., Manchester, R.~N., Lyne, A.~G., Qiao, G.~J., \& van Straten, W. 2006, \apj, 642, 868
\bibitem[Han(2009)]{Han09} Han, J.~L. 2009, in Cosmic magnetic fields: from planets, to stars and galaxies, Proc.\ IAU Symposium, ed.\ K.~G.\ Strassmeier, A.~G.\ Kosovichev, \& J.~E.\ Beckman, 259, 455
\bibitem[Harding \& Ramaty(1987)]{HR87} Harding, A.~K., \& Ramaty, R. 1987, ICRC Proc., 2, 92
\bibitem[Harding(1990)]{Harding90} Harding, A.~K. 1990, Nuclear Physics B Proceedings Supplements, 14, Issue 1, p. 3
\bibitem[Harding \& Gaisser(1990)]{HG90} Harding, A.~K., \& Gaisser, T.~K. 1990, \apj, 358, 561
\bibitem[Harding et al.(2002)]{HMZ02} Harding, A.~K., Muslimov, A.~G., Zhang, B. 2002, \apj, 576, 366
\bibitem[Harding et al.(2005)]{HUM05} Harding, A.~K., Usov, V.~V., \& Muslimov, A.~G. 2005, \apj, 622, 531
\bibitem[Harding \& Muslimov(2011a)]{HM11a} Harding, A.~K. \& Muslimov, A.~G. 2011a, \apjl, 726, L10
\bibitem[Harding \& Muslimov(2011b)]{HM11b} Harding, A.~K. \& Muslimov, A.~G. 2011b, \apj, 743, 181
\bibitem[Hessels et al.(2011)]{Hessels11} Hessels, J.~W.~T. et al. 2011, AIP Conf.\ Ser., ed.\ M. Burgay, M., N.\ D'Amico, P.\ Esposito, A.\ Pellizzoni, \& A.\ Possenti, 1357, 40
\bibitem[Hobbs et al.(2005)]{Hobbs05} Hobbs, G., Lorimer, D.~R., Lyne, A.~G., \& Kramer, M. 2005, \mnras, 360, 974
\bibitem[Hooper et al.(2009)]{Hooper09} Hooper, D., Blasi, P., \& Dario Serpico, P. 2009, \jcap, 1, 25
\bibitem[Huang et al.(2012)]{HB12} Huang, R.~H.~H. et al. 2012, \apj, 760, 92
\bibitem[Jaffe et al.(2010)]{Jaffe10} Jaffe, T.~R., Leahy, J.~P., Banday, A.~J., Leach, S.~M., Lowe, S.~R., \& Wilkinson, A. 2010, \mnras, 401, 1013
\bibitem[Johnson et al.(2014)]{Johnson14} Johnson, T.~J. et al. 2014, \apjs, 213, 6
\bibitem[Johnston et al.(2007)]{Johnston07} Johnston, S., Kramer, M., Karastergiou, A., Hobbs, G., Ord, S., \& Wallman, J. 2007, \mnras, 381, 1625
\bibitem[Kaplan et al.(2012)]{Kaplan12} Kaplan, D.~L. et al. 2012, \apj, 753, 174
\bibitem[Kashiyama et al.(2011)]{Kashiyama11} Kashiyama, K., Ioka, K., \& Kawanaka, N. 2011, \prd, 83, 023002
\bibitem[Keith et al.(2010)]{Keith10} Keith, M.~J. et al. 2010, \mnras, 409, 619
\bibitem[Keith et al.(2012)]{Keith12} Keith, M.~J. et al. 2012, \mnras, 419, 1752
\bibitem[Kennel \& Coroniti(1984)]{KC84} Kennel, C.~F. \& Coroniti, F.~V. 1984, \apj, 283, 694
\bibitem[Kiel \& Taam(2013)]{Kiel13} Kiel, P.~D. \& Taam, R.~E. 2013, \apss, 348, 441
\bibitem[Kisaka \& Kawanaka(2012)]{Kisaka12} Kisaka, S., \& Kawanaka, N. 2012, \mnras, 421, 3543
\bibitem[Kistler \& Y\"uksel(2009)]{Kistler09} Kistler, M.~D., \& Y\"uksel, H. 2009, arXiv:0912.0264
\bibitem[Kistler et al.(2012)]{Kistler12} Kistler, M.~D., Y\"uksel, H., \& Friedland, A. 2012, arXiv:1210.8180
\bibitem[Kong et al.(2012)]{Kong12} Kong, A.~K.~H. et al. 2012, \apjl, 747, L3
\bibitem[Lamb \& Yu(2005)]{Lamb2005} Lamb, F., \& Yu, W. 2005, in Binary Radio Pulsars, ASP Conf.\ Ser., ed.\ F.~A. Rasio \& I.~H.\ Stairs, 328, 299
\bibitem[Li et al.(2012)]{Li12} Li, J., Spitkovsky, A., \& Tchekhovskoy, A. 2012, \apj, 746, 60
\bibitem[Lin et al.(2015)]{Lin15} Lin, S.-J., Yuan, Q., \& Bi, X.-J. 2015, \prd, 91, 063508
\bibitem[Linden \& Profumo(2013)]{Linden13} Linden, T., \& Profumo, S. 2013, \apj, 772, 18
\bibitem[Malyshev et al.(2009)]{Malyshev09} Malyshev, D., Cholis, I., \& Gelfand, J. 2009, \prd, 80, 063005
\bibitem[Manchester et al.(2005)]{ATNF05} Manchester, R.~N., Hobbs, G.~B., Teoh, A., Hobbs, M. 2005, \aj, 129, 1993
\bibitem[Maurin et al.(2002)]{Maurin02} Maurin, D., Taillet, R., \& Donato, F. 2002, \aap, 394, 1039
\bibitem[Maurin et al.(2014)]{Maurin14} Maurin, D., Melot, F., \& Taillet, R. 2014, \aap, 569, A32
\bibitem[Moskalenko \& Strong(1998)]{Moskalenko98} Moskalenko, I.~V., \& Strong, A.~W. 1998, \apj, 493, 694
\bibitem[Moskalenko et al.(2006)]{Moskalenko06} Moskalenko, I.~V., Porter, T.~A., \& Strong, A.~W. 2006, \apj, 640, L155
\bibitem[Moskalenko(2013)]{Moskalenko13} Moskalenko, I.~V. 2013, Nuclear Phys.\ B Proc.\ Suppl., 243, 85
\bibitem[Muslimov \& Harding (2004)]{MH04} Muslimov, A. \& Harding, A.~K. 2004, ApJ, 606, 1143
\bibitem[Orlando \& Strong(2013)]{Orlando13} Orlando, E., \& Strong, A. 2013, \mnras, 436, 2127
\bibitem[Paczy\'nski(1990)]{Pacz90} Paczy\'nski, B. 1990, \apj, 348, 485
\bibitem[Pierbattista et al.(2012)]{Pierbattista12} Pierbattista, M., Grenier, I.~A., Harding, A.~K., \& Gonthier, P.~L. 2012, \aap, 545, A42
\bibitem[Pletsch et al.(2012)]{Pletsch12} Pletsch, H.~J. et al. 2012, Science, 338, 1314
\bibitem[Porter et al.(2006)]{Porter06} Porter, T.~A., Moskalenko, I.~V., \& Strong, A.~W. 2006, \apj, 648, L29
\bibitem[Porter et al.(2008)]{Porter08} Porter, T.~A., Moskalenko, I.~V., Strong, A.~W., Orlando, E., \& Bouchet, L. 2008, \apj, 682, 400
\bibitem[Porter et al.(2011)]{Porter11} Porter, T.~A., Johnson, R.~P., \& Graham, P.~W. 2011, \araa , 49, 155
\bibitem[Profumo(2012)]{Profumo12} Profumo, S. 2012, Centr.\ Eur.\ J.\ Phys., 10, 1 
\bibitem[Ransom(2010)]{Ransom10} Ransom, S.~M. et al. 2010, in AAS/High Energy Astrophysics Division \#11, BAAS, 42, 655
\bibitem[Ray et al.(2012)]{Ray12} Ray, P.~S. et al. 2012, arXiv:1205.3089
\bibitem[Ray et al.(2014)]{Ray14} Ray, P.~S. et al. 2014, AAS Meeting Abstracts, 223, \#140.07
\bibitem[Roberts(2011)]{Roberts11} Roberts, M.~S.~E. 2011,  AIP Conf.\ Ser., ed.\ M. Burgay, M., N.\ D'Amico, P.\ Esposito, A.\ Pellizzoni, \& A.\ Possenti, 1357, 127
\bibitem[Schlickeiser \& Ruppel(2010)]{Ruppel10} Schlickeiser, R., \& Ruppel, J. 2010, New J.\ Phys., 12, 033044
\bibitem[Serpico(2012)]{Serpico12} Serpico, P.~D. 2012, Astropart.\ Phys., 39, 2
\bibitem[Shaviv et al.(2009)]{Shaviv09} Shaviv, N.~J., Nakar, E., \& Piran, T. 2009, \prl, 103, 111302
\bibitem[Shklovskii(1970)]{Shklovskii70} Shklovskii, I.~S. 1970, \sovast, 13, 562
\bibitem[Sironi \& Spitkovsky(2011a)]{Sironi11a} Sironi, L. \& Spitkovsky, A. 2011a, \apj, 726, 75
\bibitem[Sironi \& Spitkovsky(2011b)]{Sironi11b} Sironi, L. \& Spitkovsky, A. 2011b, \apj, 741, 39
\bibitem[Spitkovsky(2006)]{Spitkovsky06} Spitkovsky, A. 2006, \apj, 648, L51
\bibitem[Story et al.(2007)]{Story07} Story, S.~A., Gonthier, P.~L., \& Harding, A.~K. 2007, \apj, 671, 713 (SGH)
\bibitem[Strauss \& Potgieter(2014)]{Strauss14} Strauss, R.~D., \& {Potgieter}, M.~S. 2014, Adv.\ Space Res., 53, 1015
\bibitem[Strong \& Moskalenko(1998)]{Strong98} Strong, A.~W., \&  Moskalenko, I.~V. 1998, \apj, 509, 212
\bibitem[Sturrock(1971)]{Sturrock71} Sturrock, P.~A. 1971, \apj, 164, 529
\bibitem[Venter et al.(2009)]{Venter09} Venter, C., Harding, A.~K., \& Guillemot, L. 2009, \apj, 707, 800
\bibitem[Venter et al.(2015)]{Venter15} Venter, C., Kopp, A., Harding, A.~K., Gonthier, P.~L., \& B{\"u}sching, I. 2015, Adv.\ Space Res., 55, 1529
\bibitem[Vladimirov et al.(2011)]{Vladimirov11} Vladimirov, A.~E. et al. 2011, Computer Phys.\ Comm., 182, 1156
\bibitem[Wainscoat et al.(1992)]{Wainscoat92} Wainscoat, R.~J., Cohen, M., Volk, K., Walker, H.~J., \& Schwartz, D.~E. 1992, \apjs, 83, 111
\bibitem[Yin et al.(2013)]{Yin13} Yin, P.-F., Yu, Z.-H., Yuan, Q., Bi, X.-J. 2013, \prd, 88, 023001
\bibitem[Young et al.(2010)]{Young10} Young, M.~D.~T., Chan, L.~S., Burman, R.~R., \& Blair, D.~G. 2010, \mnras, 402, 1317
\bibitem[Y\"uksel et al.(2009)]{Yuksel09} Y{\"u}ksel, H., Kistler, M.~D., \& Stanev, T. 2009, \prl, 103, 051101
\bibitem[Zhang \& Cheng(2001)]{Zhang01} Zhang, L., \& Cheng, K.~S. 2001, \aap, 368, 1063
\bibitem[Zhang \& Cheng(2003)]{ZC03} Zhang, L., \& Cheng, K.~S. 2003, \aap, 398, 639
\end{thebibliography}
\end{document}